\documentclass[aps,prx,amsmath,amssymb,superscriptaddress,twocolumn,showpacs,longbibliography,floatfix,notitlepage]{revtex4-2}

\DeclareSymbolFontAlphabet{\amsmathbb}{AMSb}%
\DeclareSymbolFontAlphabet{\mathbb}{AMSb}
\usepackage[bbgreekl]{mathbbol}

\usepackage{graphicx}
\usepackage{dcolumn}
\usepackage[colorlinks=true,linkcolor=blue,citecolor=blue,urlcolor=blue]{hyperref}
\usepackage{multirow}
\usepackage[usenames,dvipsnames]{xcolor}
\usepackage{soul}
\usepackage{braket}
\usepackage{bm}
\usepackage{nicefrac}
\usepackage{siunitx}
\usepackage{color}
\usepackage[utf8]{inputenc}   
\usepackage{upgreek}
\usepackage{ulem}
\usepackage{float}

\newcommand{\VEC}[1]{\boldsymbol{#1}}

\begin{document}

\title{Spin Models and Cluster Multipole Method: Application to Kagome Magnets}

\author{Juba Bouaziz}
\email{jbouaziz@g.ecc.u-tokyo.ac.jp}
\affiliation{Department of Physics, University of Tokyo, Tokyo 113-0033, Japan}
\author{Takuya Nomoto}
\affiliation{Department of Physics, Tokyo Metropolitan University, Hachioji, Tokyo 192-0397, Japan}
\author{Ryotaro Arita}
\affiliation{Department of Physics, University of Tokyo, Tokyo 113-0033, Japan}
\affiliation{RIKEN Center for Emergent Matter Science (CEMS), Wako 351-0198, Japan}

\date{\today}

\begin{abstract}
We present a multi-scale computational approach that combines atomistic spin models with the cluster multipole (CMP) method. The CMP method enables a systematic and accurate generation of complex non-collinear magnetic structures using symmetry-adapted representations. The parameters of the spin model are derived from density functional theory using the magnetic force theorem, with the paramagnetic state as a reference. The energy landscape of CMP-generated structures is inspected at the model Hamiltonian level, and sets of low-energy magnetic structures are identified for each material candidate. The inclusion of relativistic antisymmetric and anisotropic pair interactions lifts partially the degeneracy among these most stable structures. To demonstrate the applicability and predictive capability of the method, we apply it to the non-collinear Mn$_3X$ and collinear Fe$_3X$ ($X$ = Ga, Ge, and Sn) kagome compounds. The computational efficiency of the method in identifying low-energy structures among multiple CMP configurations highlights its potential for high-throughput screening of complex magnets with unknown magnetic order.
\end{abstract}

\maketitle

\section{Introduction}
Magnetic crystals exhibit diverse forms of long-range order, including ferromagnetism~\cite{Heisenberg1928}, antiferromagnetism~\cite{Neel1948}, and the recently identified altermagnetism, which is characterized by zero net magnetization and momentum-dependent spin polarization alternating in reciprocal space~\cite{Libor2022}. Beyond these collinear orders, magnetic moments can form non-collinear configurations, often arising from competing interatomic magnetic interactions~\cite{hughes2007,heinze2011,grytsiuk2020}. The nature of these interactions is determined by factors such as chemical composition, dimensionality, and crystal symmetry.

Non-collinear magnetic orders manifest in various forms, ranging from commensurate triangular Néel states~\cite{nakatsuji2022} and incommensurate magnetic helices~\cite{hughes2007,Bouaziz2022,Bouaziz2024} to more complex multiple-\textit{q} states~\cite{hayami2021} and topological magnetic textures such as skyrmions and merons~\cite{nagaosa2013}. These intricate magnetic configurations strongly influence the electronic band structure, leading to topological features such as Weyl nodes~\cite{nakatsuji2022,soh2024weyl}. Signatures of these topological phenomena appear in magneto-transport properties, with a notable example being the large anomalous Hall effect observed in the non-collinear kagome antiferromagnet Mn$_3$Sn~\cite{kubler2014,nakatsuji2015,nakatsuji2022}. The absence of stray fields and the potential for high-speed processing make non-collinear antiferromagnets promising candidates for spintronics applications~\cite{baltz2018,bonbien2021}.

Given the crucial role of magnetic order in determining a material's functionality, several experimental techniques have been developed to probe it, with neutron scattering~\cite{Muhlbauer2019} and magneto-transport measurements~\cite{nagaosa2013,nakatsuji2022} being widely used. On the computational side, material-specific simulations using density functional theory (DFT) have proven valuable in predicting and corroborating experimentally observed magnetic structures. High-throughput DFT calculations have been employed to investigate the magnetic ground state of collinear magnets~\cite{xu2020high} and non-collinear antiferromagnetic materials, using experimental magnetic structures as initial guesses~\cite{xu2020high}. A recently developed approach for non-collinear antiferromagnets involves evaluating within DFT the energy of constrained magnetic structures derived from the cluster multipole (CMP) method~\cite{Huebsch2021,Nomoto2024}. The CMP approach uses magnetic and toroidal multipole expansions combined with a symmetry-adapted representation to generate an orthogonal basis set of candidate magnetic structures~\cite{Suzuki2017,Suzuki2019,Suzuki2018}.

The DFT+CMP approach provides a unified framework for investigating both commensurate~\cite{Huebsch2021,Nomoto2024} and incommensurate non-collinear magnetic orders~\cite{Yanagi2023}. However, its applicability is limited by high computational cost, convergence difficulties for certain non-collinear structures within constrained DFT, and the intrinsically low energy scale of magnetic interactions (meV)~\cite{Huebsch2021}. An alternative approach is to construct an atomistic spin model with parameters derived from DFT~\cite{Sato_2010} and compare the energies of different CMP structures at the model Hamiltonian level. The key challenge then shifts to identifying a reliable spin Hamiltonian, which should:
$(i)$ reproduce the global energy landscape of arbitrary non-collinear configurations, $(ii)$ respect the hierarchy of magnetic interactions by prioritizing the largest isotropic contributions (not limited to pair interactions), and $(iii)$ incorporate non-isotropic interactions to distinguish between CMP structures with different chiralities.

Here, we present a multi-scale computational approach and systematically apply it 
to the complex kagome magnets $Tm_3X$ family, including the non-collinear anti-ferromagnetic Mn$_3X$~\cite{nakatsuji2015,soh2024weyl,kren1970neutron} and the collinear ferromagnetic Fe$_3X$ which display large anomalous Hall effect and high curie temperatures~\cite{Sales2014,Zheng2}. 
We identify the low-energy magnetic structures through DFT calculations which determine 
the atomistic spin model parameters, providing a faster and computationally efficient means 
to explore the full energy landscape of CMP structures. The parameters are derived from 
first principles using the magnetic force theorem~\cite{oguchi1983,Liechtenstein1984,Turek2006,Sato_2010,Udvardi2003,Ebert2009} in 
combination with the full-potential all-electron Korringa-Kohn-Rostoker (KKR) Green function method~\cite{Papanikolaou2002,bauer2014}. To ensure that these parameters are independent of the reference state—an essential aspect for highly non-collinear magnetic systems—the mapping is performed from a paramagnetic reference state using the disordered local moment (DLM) approach~\cite{Gyorffy_1985,staunton2006,Turek2006}. The averaging over magnetic 
orientations to simulate the paramagnetic state is achieved via the coherent potential approximation~\cite{Soven_1967,Gyorffy_1985}.

The paper is structured as follows. Sec.~\ref{all_methods} outlines the employed methods, starting with the CMP approach, followed by the DLM method, and concluding with the alloy force theorem. The results are presented in Sec.~\ref{All_results}. First, we analyze the electronic structure of $Tm_3X$ in the DLM state and examine the magnetic interactions in reciprocal space within the hexagonal Brillouin zone. Next, we investigate in detail the isotropic pair and biquadratic exchange parameters in real space.  We then compute the energies of CMP structures using the generalized Heisenberg Hamiltonian, identify the low-energy configurations, and compare them with the results obtained from magnetic constraints. Finally, we incorporate relativistic interactions and demonstrate how they select specific structures among the lowest-energy ones. Sec.~\ref{All_conclusions} provides conclusions and an outlook on potential applications of the method. The appendices include detailed derivations of the exchange parameters in the ferromagnetic limit, accounting for relativistic effects, along with the complete set of CMP-generated structures.\\[-1.0cm]

\section{Methods}
\label{all_methods}
\subsection{Cluster multipole (CMP) approach}
\label{CMP_method_sec}
We provide a concise overview of the CMP approach used to systematically generate 
the magnetic structures of the $Tm_3X$ compounds. The details of this approach can be found 
in Refs.~\onlinecite{Suzuki2017,Suzuki2018,Suzuki2019}. This expansion constitutes a \textit{discretization} of the standard multipolar expansion for a classical continuous vector potential in the Coulomb gauge~\cite{Jackson1975} \textit{focusing only on the spin magnetic moments}. The magnetic $\VEC{\mathcal{M}}_{lm}$ and toroidal $\VEC{\mathcal{T}}_{lm}$ multipoles for a collections of classical magnetic moments $\VEC{m}_i$ are introduced in a virtual atomic cluster which reflects the crystallographic 
point group of the actual crystal as~\cite{Suzuki2019}:
\begin{equation}
\VEC{\mathcal{M}}_{l\gamma} = \sum_{i=1}^{N_\text{at}} \VEC{u}^{M}_{l\gamma i}\cdot\VEC{m}_{i}\,,\,
\VEC{\mathcal{T}}_{l\gamma} = \sum_{i=1}^{N_\text{at}} \VEC{u}^{T}_{l\gamma i}\cdot\VEC{m}_{i}\,.  
\label{multipoles_def_1}
\end{equation}
\begin{equation}
\begin{split}
\VEC{u}^{M}_{l\gamma i} &= \sqrt{\frac{4\pi}{2l+1}} \VEC{\nabla}\left[\mathcal{Y}^{*}_{l\gamma}(\hat{\VEC{R}}_{i})\right]\,,\\
\VEC{u}^{T}_{l\gamma i} &= \frac{1}{l+1}(\VEC{u}^{M}_{l\gamma i}\times\VEC{R}_{i})\,\,\,.    
\end{split}
\label{multipoles_def_2}
\end{equation}
$\VEC{R}_{i}$ indicates the position of the moment $\VEC{m}_{i}$. $N_\text{at}$ designates the numbers of atoms in the cluster and $l$ is the angular momentum quantum number. For a physically meaningful expansion, the multipole basis functions $\VEC{u}^{M}_{l\gamma i}$ and $\VEC{u}^{T}_{l\gamma i}$
are given in terms of the spherical harmonics $\mathcal{Y}_{l\gamma}(\hat{\VEC{R}}_{i})$ which are 
symmetrized according to irreducible representations of the crystallographic point group:
\begin{equation}
\mathcal{Y}_{l\gamma}(\hat{\VEC{R}}_{i}) = \sum_{m}c^{\gamma}_{m}Y_{lm}(\hat{\VEC{R}}_{i})\quad.
\end{equation}
$m$ is the magnetic quantum number and $Y_{lm}(\hat{\VEC{R}}_{i})$ are the complex spherical 
harmonics. The $c^{\gamma}_{m}$ coefficients are defined and tabulated in Ref.~\onlinecite{Kusunose2009}
with $1\le\gamma\le 2l+1$. 

Using Eqs.~\eqref{multipoles_def_1} and ~\eqref{multipoles_def_2}, the atomic positions $\VEC{R}_{i}$ relative to the origin of the atomic cluster are sufficient to unambiguously define $\VEC{\mathcal{M}}_{l\gamma}$ and $\VEC{\mathcal{T}}_{l\gamma}$, with $N_\text{at}$ fixed to the number of symmetry operations of the crystallographic point group. The next step is to obtain the multipoles for the \textit{real crystal}. Focusing on commensurate magnetic structures $(\VEC{q}=0)$, this mapping procedure involves identifying cluster atoms transformed by point-group symmetry operations with their corresponding real crystal atoms transformed by space-group symmetry operations~\cite{Suzuki2019}. The mapping is not unique; for example, convenient choices are provided for the $C_{4h}$ point group in crystals with space groups $P4/m$ and $P4_2/m$ in Ref.~\onlinecite{Suzuki2019}. The multipole basis functions $\VEC{u}^{M}_{l\gamma i}$ and $\VEC{u}^{T}_{l\gamma i}$ for the real crystal is a complete basis defined in terms of spherical harmonics (Eq.~\eqref{multipoles_def_2}), they are orthogonalized using Gram-Schmidt orthonormalization procedure~\cite{Suzuki2019,Huebsch2021,Huebsch2021}.

The CMP structures generated are based on the symmetry of the crystal and do not contain information on its the magnetic energy landscape. The \textit{central assumption} within this approach is that \textit{the magnetic ground-state configuration is present among the CMP structures}, or a more elaborate scheme involving \textit{combining uniformly weighted CMPs of the same order and irreducible representation}~\cite{Huebsch2021}. This assumption was corroborated by comparing the experimental magnetic space groups from MAGNDATA~\cite{Gallego2016} for 131 materials, with $90.16\%$ of them agreeing with the CMP structures. In practice, this means that the ground state configuration for the moments $\VEC{m}_{i}$ is given by the basis functions $\VEC{u}^{M}_{l\gamma i}$ or $\VEC{u}^{T}_{l\gamma i}$, and the magnetic order is described by the CMPs. The CMP structures for the $Tm_3X$ compounds studied here are generated numerically following the procedure outlined above~\cite{Suzuki2019}. For simplicity, these structures are labeled with an integer index Nmltp in Sec.~\ref{CMP_energetics}. A detailed listing of each CMP structure, including their order $(l)$ and types (magnetic or toroidal), is provided in Appendix~\ref{Full_CMP_solutions}. 

\subsection{Disordered Local Moment}
\label{sub_sec_2}
The magnetic order and electronic properties of the $Tm_3X$ 
are investigated in the paramagnetic state, which
is modeled using the disordered local moment (DLM) theory~\cite{Gyorffy_1985} 
within spin density functional theory. The later accounts for the magnetic 
transverse fluctuations at finite temperature. The DLM approach is suitable for 
systems with strong local magnetic moment such as the Fe/Mn compounds under 
investigation~\cite{Gyorffy_1985,Staunton1992}. It relies on an adiabatic separation which 
assumes that the dynamics of the local magnetic moments $\VEC{m}_i$ is slower 
than the electronic degrees of freedom, thus the magnetic order at finite temperature 
within DFT can be modeled using various magnetic configurations of the local moments
$\VEC{m}_i$~\cite{Gyorffy_1985}. The averages over the magnetic configurations are conveniently performed using the CPA in as implemented with the KKR 
Green function approach~\cite{Ebert2011}.

In the non-relativistic limit and considering the rotational symmetry of the paramagnetic 
state, the DLM calculations can be mapped on a binary pseudo alloy with equal 
“up” and “down” concentrations~\cite{Gyorffy_1985,staunton1986,akai1993local}. The CPA approach is based on the construction of an effective medium~\cite{Soven_1967}. For a binary alloy which consists of two species A$_{x}$B$_{1-x}$ ($x$ being the concentration of the A specie), the site diagonal part of the KKR structural
Green function (see Eq.~\eqref{kkr_gf_full}) for the effective medium according to the CPA 
condition reads~\cite{akai1993local}:
\begin{equation}
{\boldsymbol{G}}^c_{ii} = x\,\bar{\boldsymbol{G}}^A_{ii} + (1-x)\,\bar{\boldsymbol{G}}^B_{ii}\,. 
\end{equation}
$\boldsymbol{G}^c_{ii}$ is a matrix in spin and orbital space $(\sigma,L)$. The local Green function 
of the specie $Q$ is obtained by embedding the scattering matrix of the element $Q$ ($\boldsymbol{t}^{Q}_i$) 
into the effective medium as follows:
\begin{equation}
   \bar{\boldsymbol{G}}^Q_{ii} = \boldsymbol{G}^c_{ii}\,[1-(\boldsymbol{t}^{Q}_{i}-\boldsymbol{t}^{c}_{i})
   \,\boldsymbol{G}^c_{ii}]^{-1}\,.
\end{equation}
$\boldsymbol{t}^{c}_{i}$ is the scattering matrix of the effective medium. The central quantity of interest are the isotropic exchange interactions (pair or biquadratic) 
in the DLM state (see Sec.~\ref{sec_alloy_FTH}). These are computed via the site off-diagonal Green functions 
$\boldsymbol{G}^{QQ^{\prime}}_{ij}$ which is given in the CPA framework by the conditionally 
averaged Green functions~\cite{Akai_1998,Sato_2010,Turek2006}:
\begin{equation}
\bar{\boldsymbol{G}}^{QQ^{\prime}}_{ij} = [1-\boldsymbol{G}^c_{ii}(\boldsymbol{t}^{Q}_i-\boldsymbol{t}^{c}_i)]^{-1}
\boldsymbol{G}^{c}_{ij}[1-(\boldsymbol{t}^{Q^{\prime}}_j-\boldsymbol{t}^{c}_j)\boldsymbol{G}^{c}_{jj}]^{-1}\,. 
\label{cond_avrg_GF}
\end{equation}
This renormalization of the effective medium's structural non-local Green 
function $\boldsymbol{G}^{c}_{ij}$ is adequate for a good description of the fully paramagnetic DLM state and strong scattering limit~\cite{Turek2006,Sato_2010}.

\subsection{Alloy force theorem}
\begin{figure*}
  \centering
  \includegraphics[width=1.0\textwidth]{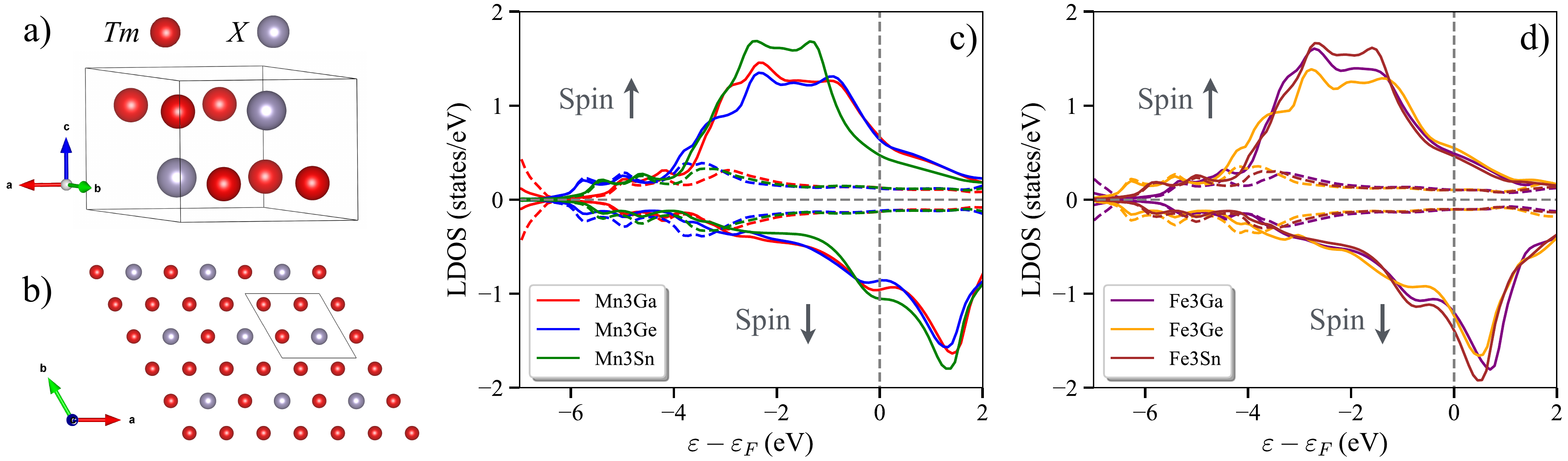}
\caption{a) Side view of the $Tm_3X$ ($Tm$ = Mn, Fe; $X$ = Ga, Ge, Sn) unit cell, showing two layers at $z = c/4$ and $z = 3c/4$. b) Top view of the $ab$ plane at $z = c/4$, where the $Tm$ atoms form a kagome lattice and the $X$ atoms form a hexagonal lattice. c-d) Spin-resolved paramagnetic local density of states (LDOS) per atom. Solid (dashed) lines represent $Tm$ ($X$) elements. The nearly filled up-spin channel and partially filled down-spin channel result in a high LDOS at the Fermi energy, indicating their metallic nature. c) Mn$_3X$ and d) Fe$_3X$ ($X$ = Ga, Ge, Sn).}
\label{geometry_Mn3X}
\end{figure*}
\label{sec_alloy_FTH}
In order to compute the magnetic interactions starting from paramagnetic state using the pseudo alloy with equal “up” and “down” concentrations~\cite{akai1993local}, i.e. the DLM state, we consider the alloy version of the magnetic force theorem~\cite{Bruno1996,Turek2006,Sato_2010}. Taking a binary alloy A$_{x}$B$_{1-x}$ ($x$ being the concentration of the A specie) with infinitesimal rotations of the local moments, the vertex corrections are cancelled to first order~\cite{Bruno1996}. Therefore, the magnetic exchange interactions can be extracted similarly to ferromagnetically ordered systems~\cite{Sato_2010}. The detailed derivation of the pair~\cite{Liechtenstein1984,Udvardi2003,Ebert2009} and biquadratic/higher order interactions~\cite{staunton1989,Mankovsky2020,Lounis2020} for the ferromagnetic state are provided in Appendix~\ref{force_theorem_FM}. For the alloy, the variation of the grand canonical potential $\delta\Omega$ is obtained from the Lloyds formula~\cite{lloyd1972,drittler1989}:
\begin{equation}
\delta\Omega = \frac{1}{\pi} \text{Im} \text{Tr}_{i\sigma Q} \int_{-\infty}^{E_{F}}
dE\ln\,(1-\bar{{\boldsymbol{\mathcal{G}}}}\delta{\boldsymbol{V}})\quad,
\label{grand_pot1}
\end{equation}
where the trace is taken over spin $(\sigma)$, sites $(i)$ and atomic species $(Q)$. The system's averaged Green 
functions $\bar{\boldsymbol{\mathcal{G}}}$ is obtained using the CPA approach (see Sec.~\ref{sub_sec_2}) and $\delta\boldsymbol{V}$ is the variation of potential due to an infinitesimal rotation of the magnetic moments. 
The analytical expression of the pair isotropic exchange  is obtained from the second order expansion (see Appendix~\ref{force_theorem_FM} and Eq.~\eqref{Append_FM_eq_1}) of $\delta\Omega$~\cite{Turek2006,Sato_2010}:
\begin{equation}
\begin{split}
{J}^{QQ^\prime}_{ij} &= 
\frac{1}{2\pi}\,\text{Im}\,\text{Tr}_{\sigma,L}
\int^{E_\text{F}}_{-\infty} dE \\&
\,\bar{\boldsymbol{G}}^{QQ^\prime}_{ij}(E) 
\,{\boldsymbol{B}}^{x,Q^\prime}_{j}(E)
\,\bar{\boldsymbol{G}}^{Q^\prime Q}_{ji}(E) 
\,{\boldsymbol{B}}^{x,Q}_{i}(E)\quad.
\end{split}
\label{jij_alloy}
\end{equation}
$\bar{\boldsymbol{G}}^{QQ^\prime}_{ij}(E)$ is the alloy structural Green function obtained from Eq.~\eqref{cond_avrg_GF}. ${\boldsymbol{B}}^{x,Q}_{i}(E)$ is the exchange splitting at site $i$, of the specie $Q$ as defined in Eq.~\ref{tmat_ebert}. The mapping is done onto a pair Heisenberg Hamiltonian for the alloy $(\mathcal{H}_\text{A})$:
\begin{equation}
\begin{split}
\mathcal{H}_\text{A} = -\sum_{i\ne j}\sum_{QQ^\prime}
{J}^{QQ^\prime}_{ij}\, 
(\VEC{e}^{Q}_{i}\cdot \VEC{e}^{Q^\prime}_{j})\quad,\\
\end{split}
\label{grand_pot4}
\end{equation}
where $\{Q,Q^\prime\}=\{A,B\}$ for the binary alloy. In the particular case of the DLM state, the following symmetry relations apply ${J}^{\uparrow\uparrow}_{ij} = {J}^{\downarrow\downarrow}_{ij}$ and ${J}^{\uparrow\downarrow}_{ij}=-{J}^{\uparrow\uparrow}_{ij}$. Therefore, $\{Q,Q^\prime\}=\{\uparrow,\uparrow\}$ is sufficient to extract the magnetic interactions in the DLM state. Eq.~\eqref{jij_alloy} is then simplified further by introducing $\bar{{\boldsymbol{G}}}_{ij}(E) = \bar{{\boldsymbol{G}}}^{\uparrow\uparrow}_{ij}(E)$:
\begin{equation}
\begin{split}
{J}_{ij} = 
\frac{1}{2\pi}\,\text{Im}\,\text{Tr}_{\sigma,L}
\int^{E_\text{F}}_{-\infty} dE 
\,\bar{{\boldsymbol{G}}}_{ij}
\,{\boldsymbol{B}}^{x,\uparrow}_{j}
\,\bar{{\boldsymbol{G}}}_{ji}
\,{\boldsymbol{B}}^{x,\uparrow}_{i}\quad.
\end{split}
\label{jij_alloy_2}
\end{equation}
The explicit energy dependencies are dropped for a more compact expression. The isotropic biquadratic interactions in the DLM state are extracted using the fourth order expansion (see Appendix~\ref{force_theorem_FM} and Eq.~\eqref{biquadratic_ferro}) of $\delta\Omega$~\cite{Mankovsky2020}:
\begin{equation}
\begin{split}
{B}_{ij} &= 
\frac{1}{2\pi}\,\text{Im}\,\text{Tr}_{\sigma,L}
\int^{E_\text{F}}_{-\infty} dE\,
[
\,\bar{{\boldsymbol{G}}}_{ij} 
\,{{\boldsymbol{B}}}^{x,\uparrow}_{j} 
\,\bar{{\boldsymbol{G}}}_{ji} 
\,{{\boldsymbol{B}}}^{x,\uparrow}_{i}]\\&
\times[
\,\bar{{\boldsymbol{G}}}_{ij}  
\,{{\boldsymbol{B}}}^{x,\uparrow}_{j} 
\,\bar{{\boldsymbol{G}}}_{ji} 
\,{{\boldsymbol{B}}}^{x,\uparrow}_{i}]
\quad.
\end{split}
\label{biq_is_formula}
\end{equation}
The biquadratic interactions obtained from Eq.~\eqref{biq_is_formula} for elemental itinerant magnets such as bcc Fe and fcc/hcp Co are found to be in very good agreement with the ones obtained from the DLM based spin cluster expansion technique~\cite{Szunyogh2011,hatanaka2024}. The Heisenberg model Hamiltonian including isotropic non-relativistic pair and biquadratic interactions reads:
\begin{equation}
\begin{split}
\mathcal{H} &= -\sum_{ij} J_{ij}\,(\VEC{e}_{i}\cdot\VEC{e}_{j})
-\sum_{ij} B_{ij}\,(\VEC{e}_{i}\cdot\VEC{e}_{j})^{2}\quad.
\end{split}
\label{gen_heis_kkr_iso}
\end{equation}
$\VEC{e}_{i}$ being a the unit vector for the classical magnetic moments $\VEC{m}_i$ introduced in Sec.~\ref{CMP_method_sec}. The inclusion of relativistic spin-orbit interactions introduces more complex terms to the Hamiltonian including anti-symmetric terms and symmetric anisotropic terms~\cite{Ebert2009,bouaziz2017}. The anti-symmetric Dzyaloshinskii–Moriya interaction~\cite{Dzyaloshinsky1964,Moriya1960} in the Mn$_3X$ compounds is addressed in Sec.~\ref{DM_Mn3X}. 

\section{Results and discussion}
\label{All_results}
\subsection{Crystal and electronic structure}
\label{setup_details}
\begin{figure*}
  \centering
  \includegraphics[width=0.8\textwidth]{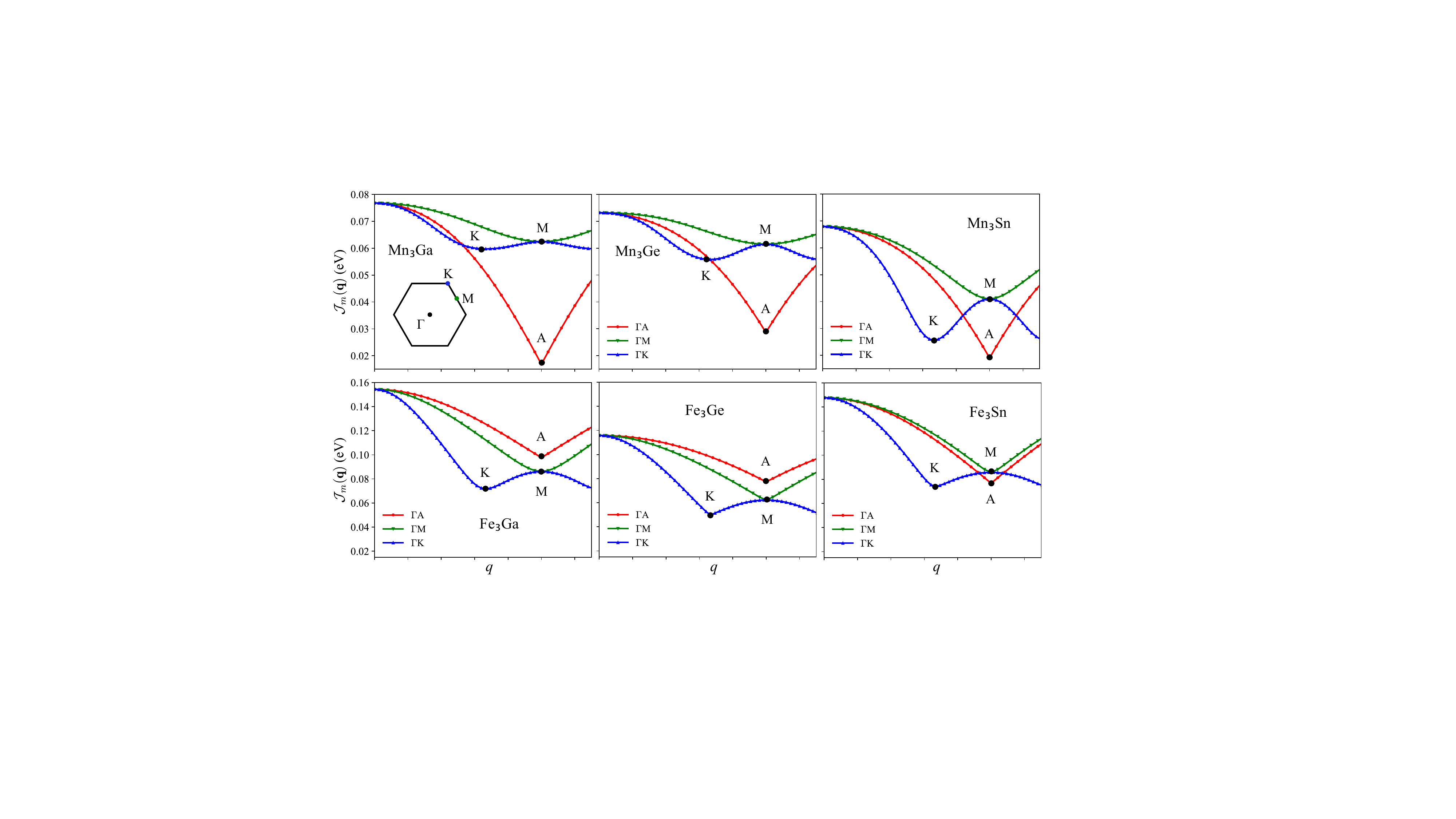}
\caption{The maximal eigenvalue, $\mathcal{J}_\text{m}(\VEC{q})$, is shown for all $Tm_3X$ 
compounds. Three high-symmetry lines in the hexagonal Brillouin zone are shown: 
$\Gamma$A, $\Gamma$M, and $\Gamma$K. The inset in the first panel illustrates 
the two-dimensional Brillouin zone, marking the (M, K) points, while the A point 
lies along the $c$-direction. The $x$-axis has been rescaled to better display 
the dispersion and symmetry of $\mathcal{J}_\text{m}(\VEC{q})$.}
\label{Jq_Tm3X}
\end{figure*}
The first-principles calculations were performed using the all-electron full-potential 
KKR Green function method~\cite{juKKR:22} within the scalar 
relativistic approximation, employing $l_\text{max} = 3$ for the orbital expansion of 
the Green function. The self-consistent calculations are performed using a k-mesh of 
$25\times25\times25$, an energy contour containing 58 energy points and an electronic 
smearing temperature of $500$ K. The paramagnetic electronic state electronic structure 
is obtained using the charge self-consistent DLM approach~\cite{Turek2006}. The magnetic exchange 
interactions are extracted using the alloy force theorem (see Sec.~\ref{sec_alloy_FTH}) using 
a denser k-mesh $30\times30\times30$ and an electronic temperature of $300$ K. 

We focus on the $Tm_3X$ ($Tm$ = Mn, Fe; $X$ = Ga, Ge, Sn) compounds in their DO$_{19}$ structure, which belongs to the hexagonal space group P63/mmc. The unit cell, shown in Fig.~\ref{geometry_Mn3X}a, contains eight atoms (six $Tm$ and two $X$ atoms). The $Tm$ atoms occupy two distinct  layers ($z = c/4$ 
and $z = 3c/4$), each layer forming a two-dimensional kagome lattice (see 
Fig.~\ref{geometry_Mn3X}b). The hexagonal lattice constants $(a, c)$ are taken from 
experiment and given in Table~\ref{lattice_magn_1}. 

The paramagnetic local density of states (LDOS) for one representative $Tm$ and $X$ atom of $Tm_3X$ is shown 
in Figs.~\ref{geometry_Mn3X}c-d. It consists of the imaginary part of the local Green function for the $\uparrow$-component of the pseudo-alloy. The electronic structure is broadened due to the thermal magnetic fluctuation~\cite{dlm}. The local magnetic moment 
(spin-splitting) is present on the $Tm$ sites while the $X$ atoms are non-magnetic in contrast to the ferromagnetic 
state where the $spd$-orbitals of $X$ acquire an induced spin polarization. The high density of states at the Fermi 
energy indicates that systems are metallic and the $Tm$ magnetic moments likely interact via an indirect Ruderman–Kittel–Kasuya–Yosida (RKKY) exchange mechanism~\cite{park2018}. Note that for the Mn$_3$Ga and Fe$_3$Ga compounds, the Ga $3d$-states are visible at around $\varepsilon\simeq -7$ eV away from the Fermi energy. The localization of the magnetic moment on the $Tm$ sites makes 
a convenient state to map onto a Heisenberg model~\cite{Gyorffy_1985}. The comparison between 
the Mn and Fe LDOS shows that the additional electron in the  Fe $3d$ spin down channel systematically lead to a 
reduction the spin moment of Fe$_3X$  with respect to Mn$_3X$ as shown in Table~\ref{lattice_magn_1}, which is also 
in agreement  with Hund's rules independently of the non-magnetic $X$-elements. For both Mn$_3X$ and Fe$_3X$, the largest magnetic moments are obtained for $X$ = Sn, intermediate moments for $X$ = Ga, and the smallest for $X$ = Ge (see Table~\ref{lattice_magn_1}). The calculated magnetic moments on the $Tm$ atoms are compared with experimental values, which were measured at different temperatures and include error bars. Overall, the results show reasonable agreement: the moments of Mn$_3X$ and Fe$_3$Ga are slightly overestimated, while those of Fe$_3$Ge and Fe$_3$Sn are underestimated.

\begin{table}
\centering
\begin{tabular}{ccccc}
\hline
$Tm_3X$  & a $(\AA)$  &  c $(\AA)$  & $M^\text{exp}(\mu_\text{B})$  & $M_\text{DLM}(\mu_\text{B})$\\
\hline
Mn$_3$Ga\cite{kren1970neutron}  & 5.360 & 4.325 & 2.40                   & 2.76 \\
Mn$_3$Ge\cite{kadar1971neutron} & 5.360 & 4.320 & 2.50\cite{Soh2020}    & 2.65 \\
Mn$_3$Sn\cite{cable1993}        & 5.665 & 4.531 & 3.17                   & 3.22 \\
Fe$_3$Ga\cite{Couderg1971}      & 5.200 & 4.360 & 2.03\cite{Zezhong2024} & 2.24 \\
Fe$_3$Ge\cite{Andrusyak1991}    & 5.178 & 4.226 & 2.20\cite{Zheng2}      & 2.04 \\
Fe$_3$Sn\cite{Buschow1983}      & 5.457 & 4.362 & 2.37\cite{Sales2014}   & 2.27 \\ 
\hline
\end{tabular}
\caption{Experimental lattice constant parameters for the studied $Tm_3X$ compounds. 
The magnetic moments obtained from experiment (M$^\text{exp}$) and the spin moments 
from the present calculations (M$_\text{DLM}$) are shown in the last two rows.}
\label{lattice_magn_1}
\end{table}

\subsection{Magnetic interactions in reciprocal space}
\label{Jij_reciprocal}
The $Tm_3X$ family has different competing magnetic interactions at play which give it its rich and 
complex magnetic phase diagram. Here, we focus on the isotropic magnetic interactions $J_{ij}$, computed 
from the DLM state which determine the transition temperature ($T_\text{p}$) from a 
magnetically ordered state to a paramagnetic one, and the possible emergence of incommensurate
magnetic order below $T_\text{p}$ due to frustrated exchange interactions~\cite{Bouaziz2022,Bouaziz2023}. 
Considering the translation symmetry of $Tm_3X$ containing six sites per unit cell, 
it is convenient to split the site indices $\{i,j\}$ introduced in Eq.~\eqref{gen_heis_kkr_iso} into 
$i=\{m,k\}$, where $m$ indicates the atomic index in the unit cell and $k$ 
is the cell index with the position $\VEC{R}^{k}_{m}$. The Fourier transform of the isotropic pair interactions reads:
\begin{equation}
{\mathcal{J}}_{mn}(\VEC{q}) = \sum_{k=1}^{N_c}e^{i\VEC{q}\cdot(\VEC{R}^{k}_{n}-\VEC{R}^{0}_{m})}{J}^{0k}_{mn}\quad. 
\label{Fourier_tf}
\end{equation}
$N_c$ is the number of unit cells, and $\underline{\mathcal{J}}(\VEC{q})$ 
is a $6\times6$ matrix. The Fourier transforms are computed with a real-space cutoff 
$R_\text{cut}=2a$ (where $a$ is the in-plane lattice constant), a distance at which 
all $J_{ij}$ for $Tm_3X$ vanish as shown in Fig.~\ref{Jij_real_analysis}. We now inspect the dispersion along 
three high-symmetry directions within the hexagonal Brillouin zone of the highest eigen 
value of $\underline{\mathcal{J}}(\VEC{q})$, $\mathcal{J}_\text{m}(\VEC{q})$. 
The results are shown in Fig.~\ref{Jq_Tm3X}, the wave vector $\VEC{q}_\text{m}$ that maximizes 
$\mathcal{J}_\text{m}(\VEC{q})$ determines the lowest-energy magnetic modulation~\cite{Mendive2019,Bouaziz2022,Bouaziz2024}. 
For all $Tm_3X$ compounds, $\mathcal{J}_\text{m}(\VEC{q})$ reaches its maximum at $\VEC{q}_m=0$, 
indicating that the magnetic structures are commensurate between crystallographic cells. 
For Mn$_3X$ compounds, these results are in good agreement with Ref.~\onlinecite{Mendive2019}, 
where it was found that magneto-elastic effects do not lead to the formation of incommensurate magnetic 
order. The magnetic structure within each cell will be analyzed in in details in Sec.~\ref{CMP_energetics} 
using the CMP-generated magnetic structures. We note that, for Mn$_3$Sn in the 
low-temperature regime, an incommensurate helical magnetic order emerges along the $c$-direction~\cite{cable1993,duan2015,park2018,Chen2024}, 
characterized by a wave vector $q_{z} = 0.09\,(2\pi/c)$. This incommensurate ordering is 
reproduced when the exchange parameters are derived from the ferromagnetic reference state and 
including Hubbard $(U,J)$ to the DFT exchange-correlation potential~\cite{park2018}.
\begin{table}
\centering
\begin{tabular}{cccc}
\hline
Tm$_3$X  & T$^\text{th}_{p}$ (K) & T$^\text{exp}_{p}$ (K)\\
\hline
Mn$_3$Ga &  594.2 & 470~\cite{kren1970neutron}  \\
Mn$_3$Ge &  566.1 & 380~\cite{Soh2020}     \\
Mn$_3$Sn &  525.8 & 420~\cite{zimmer1972}  \\
Fe$_3$Ga & 1194.1 & 720~\cite{Zezhong2024} \\
Fe$_3$Ge &  896.1 & 640~\cite{Zheng2}      \\
Fe$_3$Sn & 1142.8 & 725~\cite{Sales2014}   \\
\hline
\end{tabular}
\caption{Comparison between the theoretical and experimental transition temperatures of $Tm_3X$ in Kelvin (K).
The theoretical values $T_p$ are obtained using the mean-field approximation (using Eq.~\ref{Tc_meanF}), while the experimental values $T^\text{exp}_p$ are taken from previous measurements.}
\label{Tc_all}
\end{table}

At the mean-field level, the key quantity for determining the transition temperature is 
$\mathcal{J}_\text{m}(\VEC{q}_\text{m})$, and $T_\text{p}$ reads~\cite{Turek2006}:
\begin{equation}
T_\text{p} = \frac{2}{3k_\text{b}}\mathcal{J}_\text{m}(0)\quad,
\label{Tc_meanF}
\end{equation}
where $k_\text{b}$ is the Boltzmann constant. The obtained $T_\text{p}$ values for all 
$Tm_3X$ compounds are provided in Table~\ref{Tc_all}, which also includes the corresponding 
experimental values. The calculated $T_\text{p}$ values are systematically higher than 
the experimental ones, as expected from a mean-field approximation~\cite{Mendive2019}. 
The experimental trends are well reproduced for the Mn$_3$Ga, Mn$_3$Sn, and Fe$_3X$ 
compounds, with the exception of the Mn$_3$Ge system, where the computed $T_\text{p}$ 
value exceeds that of Mn$_3$Sn.

\subsection{Magnetic interactions in real space: pair exchange}
\begin{figure}[H]
  \centering
  \includegraphics[width=0.42\textwidth]{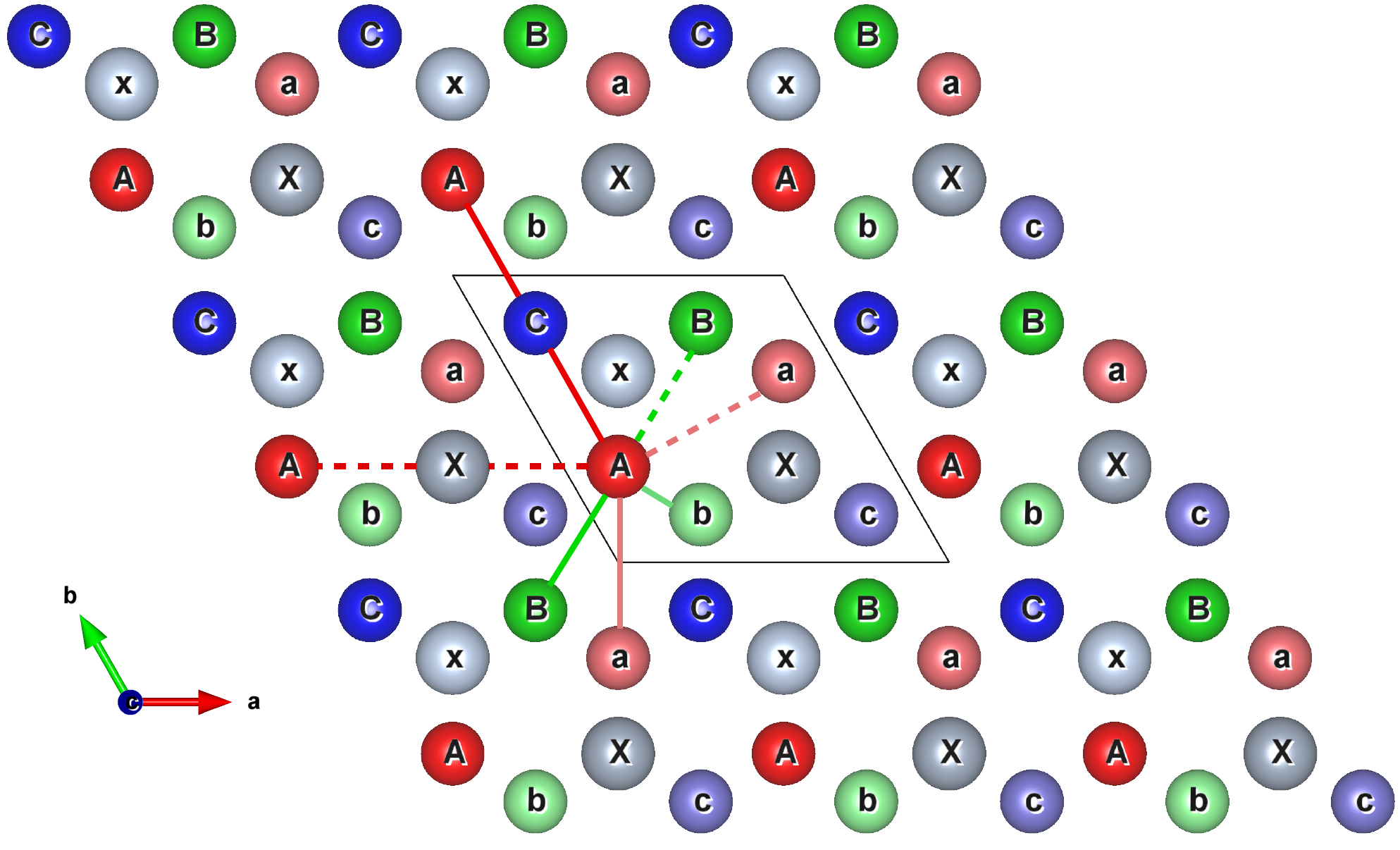}
\caption{The top view of the unit cell shows the $Tm$ sites at $c=z/4$, denoted by capital 
letters A, B, and C, while the non-magnetic site is labeled $X$. The sites at $c=3z/4$ are 
represented by lowercase letters a, b, c, and $x$. The primitive cell is outlined by the 
black rhombus. The bonds indicate the four representative interactions analyzed. Inequivalent 
interactions at equal distances are depicted using solid and dashed lines.}
\label{sub_jij_fig}
\end{figure}

We now examine the real-space dependence of the isotropic pair magnetic interactions $J_{ij}$. The $Tm_3X$ compounds contain six atoms per unit cell, so the full exchange 
matrix has a size of $6\times6$ and a cell dependence (${J}^{kl}_{mn}$ introduced in Sec.~\ref{Jij_reciprocal}). The components of this matrix are determined using 
Eq.~\eqref{jij_alloy_2}. The different sites within each plane 
($z=c/4$ and $z=3c/4$) are related by $C_{3}$ symmetry, while the atoms from one kagome 
layer are related to their partners in the other layer via inversion symmetry, e.g. atom A 
in Fig.~\ref{sub_jij_fig} transforms into $a$. The analysis of couplings between the different 
sublattices in Fig.~\ref{sub_jij_fig} allows the reduction of the full exchange matrix to 
a group of four representative interactions.

\begin{figure*}
  \centering
  \includegraphics[width=1.0\textwidth]{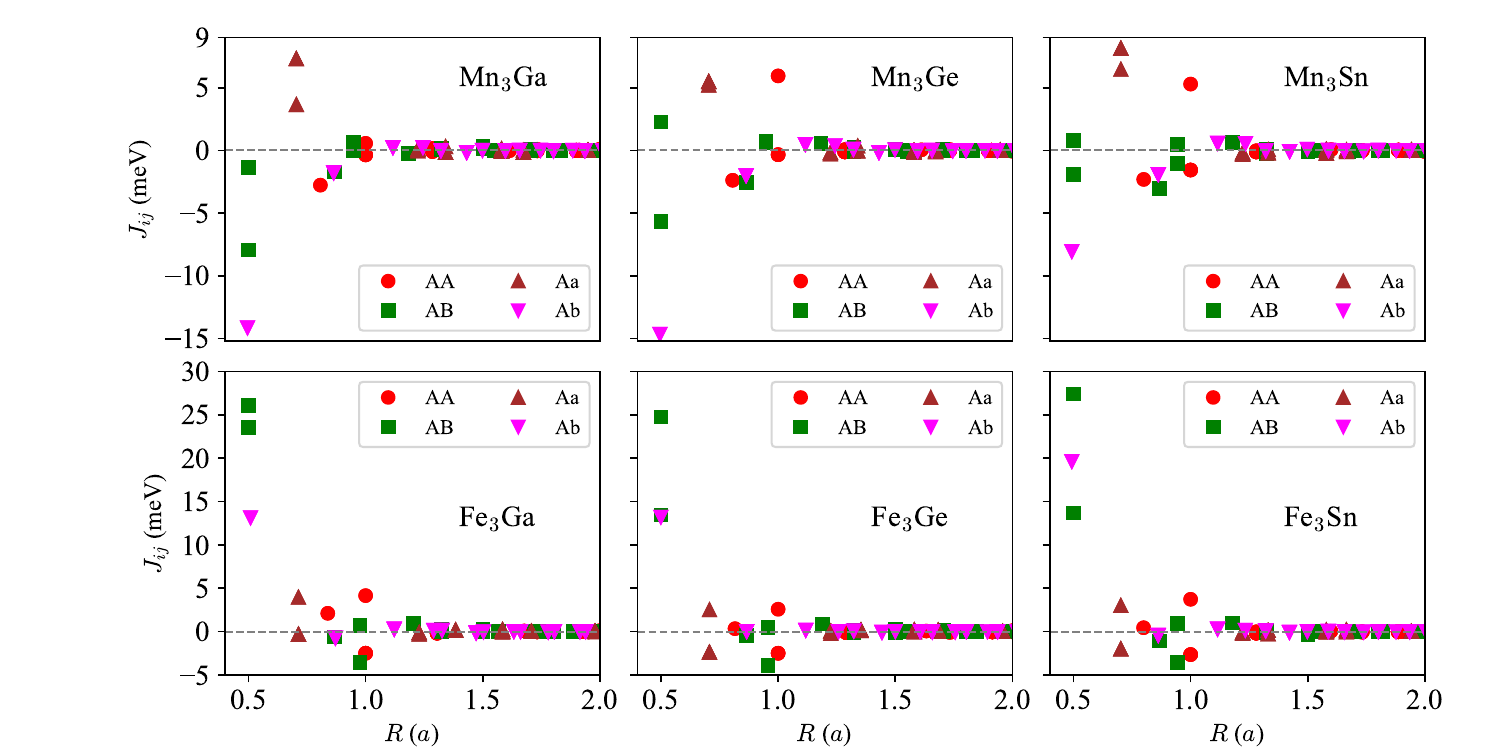}
\caption{Isotropic pair interactions $J_{ij}$ as a function of the interatomic distance $R$ 
(given in units of the in plane lattice constant $a$) for the representative sublattice couplings:
AA, Aa, AB, and Ab, indicated in different colors. Each panel corresponds to one of the $Tm_3X$ 
compounds ($Tm$ = Mn, Fe; $X$ = Ga, Ge, Sn). The in plane exchange interactions in Mn$_3X$ favors 
antiferromagnetism, while in Fe$_3X$, it tends to be ferromagnetic.}
\label{Jij_real_analysis}
\end{figure*}

The magnetic interactions for all the $Tm_3X$ compounds are shown in Fig.~\ref{Jij_real_analysis}. 
Each panel contains four groups: AA, AB, Aa, and Ab. Within our spin Hamiltonian convention, positive 
interactions indicate ferromagnetic coupling (FM), while negative interactions signify antiferromagnetic 
coupling (AFM). The interactions exhibit oscillatory negative and positive values up to a range of 1.5 $a$ 
($a$ being the in-plane lattice constant), which do not necessarily follow an RKKY exchange decay. 
This behavior aligns with previous theoretical and experimental findings for Mn$_3$Sn in Ref.\cite{park2018}, 
where it was suggested that the discrepancy may arise either from the non-sphericity of the Fermi surface~\cite{Chen2024} 
or from multi-band contributions~\cite{akbari2013}. A common feature observed across all $Tm_3X$ compounds 
is the inequivalence of the first nearest-neighbor (NN) interactions for AB coupling. This arises because the 
connecting bond either includes an $X$ or Mn atom on the top left (Fig.~\ref{sub_jij_fig}), 
leading to significantly different AB interactions, which may even be opposite in nature, as observed in 
Mn$_3$Ge and Mn$_3$Sn~\cite{Nyari2019}. Similarly, inequivalence is evident in the interlayer couplings Aa, 
where the connecting bond involves either two $X$ atoms or two Tm atoms (Fig.~\ref{sub_jij_fig}). Lastly, 
the AA coupling within the same plane is non-negligible and exhibits a substantial splitting at similar 
distances for Mn$_3$Ge, Mn$_3$Sn, and all Fe$_3X$ compounds. This difference arises from whether the 
connecting bond passes through a $Tm$ atom or an $X$ atom (Fig.~\ref{sub_jij_fig}).

The Mn compounds shown in the first row of Fig.~\ref{Jij_real_analysis} exhibit interlayer nearest-neighbor 
(NN) Ab coupling that favors an antiferromagnetic (AFM) order. Among them, Mn$_3$Ge has the largest values, 
while Mn$_3$Sn exhibits the smallest. The NN interlayer Aa coupling is systematically ferromagnetic (FM), 
whereas the NN interaction for AA coupling is AFM. These AA couplings, which represent interactions between 
atom A and another atom A in the adjacent unit cell along the $c$-direction, are not displayed in Fig.~\ref{sub_jij_fig}. 
The Fe$_3X$ compounds under consideration exhibit strong FM interactions for both AB and Ab couplings. 
AFM interactions are observed for the AA coupling between next-nearest neighbors (in-plane), for third 
neighbors in the AB coupling, and for NN in the Aa coupling. Despite the presence of these AFM interactions, 
the strong NN FM interactions for Ab and AB couplings dominate in magnitude. Overall, the comparison between 
Mn$_3X$ and Fe$_3X$ compounds reveals that the former tends toward AFM orders, while the latter favors 
FM orders.

\subsection{Magnetic interactions in real space: biquadratic exchange}
\begin{figure*}
  \centering
  \includegraphics[width=1.0\textwidth]{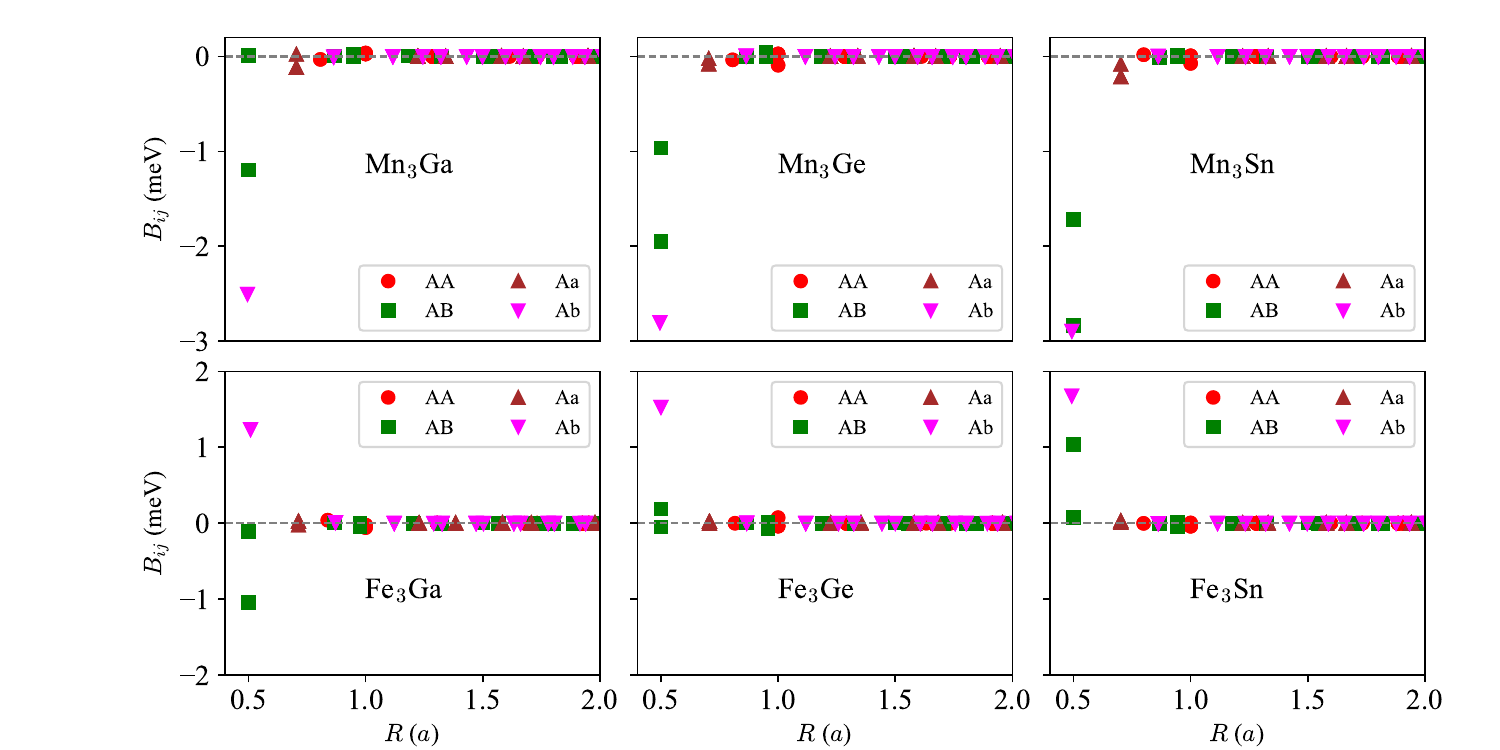}
\caption{Isotropic biquadratic interactions $B_{ij}$ as a function of the interatomic distance $R$ 
(given in units of the in plane lattice constant $a$) for the representative sublattice couplings: 
AA, Aa, AB, and Ab, indicated in different colors. Each panel corresponds to one of the $Tm_3X$ compounds 
($Tm$ = Mn, Fe; $X$ = Ga, Ge, Sn). The $B_{ij}$ interactions have a shorter spatial range compared to the 
$J_{ij}$ interactions and promote a perpendicular alignment of the magnetic moments in the Mn$_3X$ 
compounds.}
\label{Bij_real_analysis}
\end{figure*}

Having analyzed the pair isotropic interactions, we now turn to the biquadratic isotropic interactions, $B_{ij}$, 
which represent the next largest isotropic contributions. Within our spin Hamiltonian convention 
(see Eq.~\eqref{gen_heis_kkr_iso}), a positive $B_{ij}$ coefficient favors collinear alignment of magnetic moments, 
whereas a negative $B_{ij}$ promotes perpendicular alignment. The biquadratic interactions share the same symmetry 
properties as the pair isotropic interactions with respect to the real-space indices $(i,j)$. Consequently, similar 
to the $J_{ij}$ parameters, the $B_{ij}$ interactions will be discussed in terms of the representative groups AA, AB, Aa, and Ab, 
as depicted in Figure~\ref{Bij_real_analysis}. The range of the $B_{ij}$ coefficients is significantly shorter than 
that of the $J_{ij}$ Beyond the first nearest neighbor (NN) for each group, these interactions become negligibly small. 
This behavior is consistent with the analytical expression for $B_{ij}$ (Eq.~\eqref{biq_is_formula}), which involves 
a fourth-order product of non-local Green's functions that decay with increasing interatomic distance. 
In contrast, the $J_{ij}$ interactions depend on a second-order product of Green's functions (Eq.~\eqref{jij_alloy}), 
resulting in a slower decay. A similarly rapid decay of the $B_{ij}$ parameters has been calculated for bcc Fe 
iron~\cite{Mankovsky2020, hatanaka2024} and Ru-based Heusler alloys~\cite{Simon2020}. 

For all Mn compounds, the $B_{ij}$ coefficients favor a perpendicular alignment. In the $Ab$ group, the NN $B_{ij}$ 
values are $~0.16 J_{ij}$ for Mn$_3$Ga and Mn$_3$Ge, and $0.35 J_{ij}$ for Mn$_3$Sn. For the $AB$ group, the $B_{ij}$ 
coefficients at the same distance exhibit a splitting analogous to the $J_{ij}$, arising from the inequivalence of 
the bonds, as previously discussed (see Fig.~\ref{sub_jij_fig}). The trend remains consistent: weaker $J_{ij}$ 
correspond to weaker $B_{ij}$. On one hand, the largest AB biquadratic interaction for Mn$_3$Ga is $0.15 J_{ij}$, 
while for Mn$_3$Ge, it is $0.34 J_{ij}$. On the other hand, $B_{ij}$ for Mn$_3$Sn exceeds the corresponding $J_{ij}$, 
reaching $1.47 J_{ij}$. The interlayer $Aa$ biquadratic interactions are negative and relatively weak compared to 
the $Ab$ and $AB$ interactions. For the Fe compounds, the largest $B_{ij}$ coefficients are again associated with 
the Ab coupling, favoring a collinear alignment. In Fe$_3$Ga, the AB couplings promote a perpendicular alignment, 
whereas they favor a collinear configuration in Fe$_3$Ge and Fe$_3$Sn. Additionally, the $B_{ij}$ coefficients for 
the Aa couplings are negligibly small. Overall, the largest $B_{ij}$ values are obtained for nearest AB and Ab couplings for all the $Tm_3X$ compounds. 

The significance of higher-order terms in Mn$_3X$ compounds has already been highlighted in Refs.~\onlinecite{Mendive2019,Dias2021}, 
which used a Weiss field-based DLM approach~\cite{Gyorffy_1985} and total energy fits to extract effective spin model 
parameters, respectively. These studies also identified isotropic multi-site interactions, such as three-site 
interactions between Mn triangles. Such interactions, however, are not included in the present study and will be addressed in future work. 

\subsection{Cluster multipole energies}
Considering the DLM reference state magnetic interactions, all the $Tm_3X$ systems under investigation exhibit a 
commensurate magnetic order, i.e., Fourier transforms peak at $\VEC{q}_\text{m}=0$, as shown in Sec.~\ref{Jij_reciprocal}. 
Therefore, the magnetic Hamiltonian given in Eq.~\eqref{gen_heis_kkr_iso} will be restricted to these 
commensurate structures and recast as in Eq.~\eqref{Fourier_tf} using sublattice (${m,n}$) and cell $({k,l})$ 
indices~\cite{Nyari2019}:
\begin{equation}
\begin{split}
\mathcal{H} &= -\sum_{kl, mn} J^{kl}_{mn}\,(\VEC{e}^{k}_{m}\cdot\VEC{e}^{l}_{n})
-\sum_{kl, mn} B^{kl}_{mn}\,(\VEC{e}^{k}_{m}\cdot\VEC{e}^{l}_{n})^2\quad,
\end{split}
\label{gen_heis_kkr}
\end{equation}
and the Hamiltonian per unit cell is then given by:
\begin{equation}
\begin{split}
\mathcal{H}_\text{c} & = -\sum_{mn} \mathcal{J}_{mn}(\VEC{e}_{m}\cdot\VEC{e}_{n})
-\sum_{mn}\mathcal{B}_{mn}(\VEC{e}_{m}\cdot\VEC{e}_{n})^2\quad.\\
\end{split}
\label{sub_heisenberg}
\end{equation}
The sublattice interactions are defined as $\mathcal{J}_{mn} = \sum_{l} J^{0l}_{mn}$, and 
$\VEC{e}_{n}$ represents the unit vector of cell $0$ for sublattice $n$. The summation ${m,n}$ 
in Eq.~\eqref{sub_heisenberg} runs over all sublattices $\{$A, B, C, a, b, c$\}$ in the $Tm_3X$ unit cell. 
The values of the sublattice pair ($\mathcal{J}_{mn}$) and biquadratic ($\mathcal{B}_{mn}$) 
interactions are provided in Table~\ref{Tab_all_sub_int}. 
\label{CMP_energetics}
\begin{figure*}
\centering
 \includegraphics[width=1.0\textwidth]{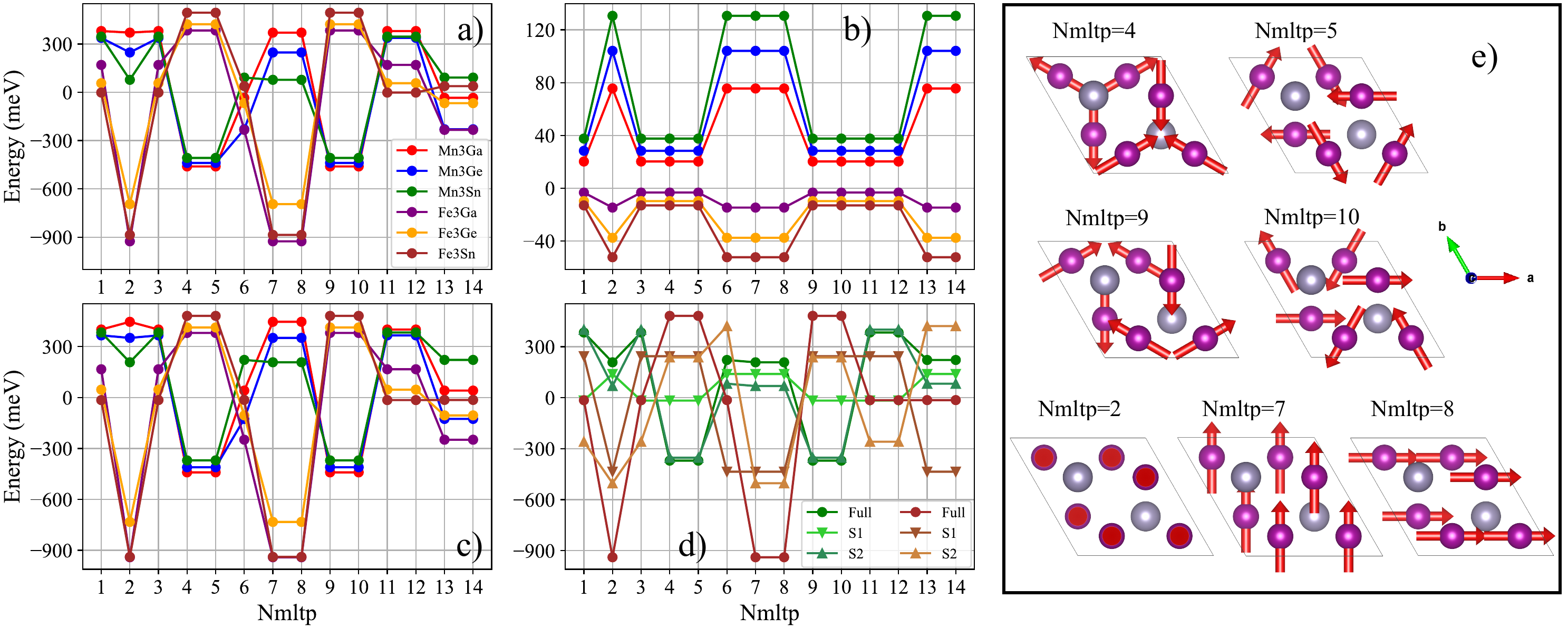}
  \caption{a) Magnetic energy per unit cell as function of the CMP solutions considering only isotropic pair exchange 
  interactions $J_{ij}$. b) Same as in (a) but considering only isotropic biquadratic exchange interactions $B_{ij}$. 
  c) Same as in (a) but considering isotropic pair and biquadratic exchange interactions $J_{ij}$ and $B_{ij}$. 
  d) Magnetic energy per unit for Mn$_3$Sn (green curves) and Fe$_3$Sn (brown curves) including $J_{ij}$ and $B_{ij}$. 
  Full: Indicates energy curves that include all sublattice interactions. S1: Indicates energy curves that include only 
  interactions within the sublattices $\{$A, B, C$\}$ or $\{$a, b, c$\}$. S2: Indicates energy curves that include only 
  interactions from $\{$A, B, C$\}$ to $\{$a, b, c$\}$. The indexation of the sublattices follows the one given in 
  Fig.~\ref{sub_jij_fig}. e) The minimal energy CMP solutions which features: $4$ antiferromagnetic non-collinear 
  solutions for the Mn$_3X$ compounds and $3$ collinear solutions for the Fe$_3X$ ones. }
\label{Fig_CMP_energies}
\end{figure*}

We now systematically compare the magnetic energy per unit cell using Eq.~\eqref{sub_heisenberg}, 
considering multiple commensurate collinear and non-collinear structures generated using the CMP 
method (see Sec.~\ref{CMP_method_sec}). The CMP generates a total of $14$ structures for which the moments $\VEC{m}_{i}$ have fixed magnitude, all of which are presented in Appendix~\ref{Full_CMP_solutions}. Fig.~\ref{Fig_CMP_energies} a-d show the energies as a function of the Nmltp (multipole index). Several CMP solutions are degenerate, as only isotropic interactions are included. At this point, the aim is to determine, out of all the possible CMP structures, lowest energy sets by considering the largest isotropic magnetic interactions (pair and biquadratic).
\begin{table}
\centering
\begin{tabular}{ccccccccc}
\hline
$Tm_3X$  & $\mathcal{J}_{AA}$ & $\mathcal{J}_{AB}$ & $\mathcal{J}_{Aa}$  & $\mathcal{J}_{Ab}$ & $\mathcal{B}_{AA}$ & $\mathcal{B}_{AB}$ & $\mathcal{B}_{Aa}$ & $\mathcal{B}_{Ab}$\\
\hline
Mn$_3$Ga & -4.89 & -11.56 & 35.5 & -34.62 & 0.12 & -1.12 & -0.42 & -5.03\\
Mn$_3$Ge & 5.13 & -3.32 & 29.88  & -34.84 & -0.15 & -2.83 & -0.37 & -5.58\\
Mn$_3$Sn & -1.24 & -6.41 & 42.25 & -20.56 & -0.08 & -4.55 & -1.03 & -5.77\\
Fe$_3$Ga & 1.44 & 47.64 & 7.34   & 25.18  & -0.14 & -1.25 & 0.05 & 2.51\\
Fe$_3$Ge & -5.41 & 34.5 & -2.91  & 27.61  & -0.01 & 0.02 & 0.09 & 3.07\\
Fe$_3$Sn & -3.79 & 37.23 & -1.87 & 39.46  & -0.15 & 1.02 & 0.14 & 3.34\\
\hline
\end{tabular}
\caption{Representative sublattice interaction parameters for $Tm_3X$ compounds, including 
isotropic pair exchange and biquadratic exchange interactions. All coefficients are provided 
in units of meV.}
\label{Tab_all_sub_int}
\end{table}

Fig.~\ref{Fig_CMP_energies}a shows the energy landscape including $\mathcal{J}_{mn}$ terms. 
These split into two separate groups: one for Mn$_3X$ and one for Fe$_3X$. The first group, Mn$_3X$, 
has four non-collinear degenerate lowest energy solutions, which are depicted in Fig.~\ref{Fig_CMP_energies}e 
(Nmltp=4,5,9,10). These results are in agreement with previous findings from the literature~\cite{sticht1989,Nyari2019}. 
The common feature among all these solutions is that they present a triangular non-collinear order with a 
$120^{\circ}$ relative angle within each Kagome layer, stabilized by the AFM $\mathcal{J}_{AB}$ coupling. 
The moments on each sublattice are then coupled ferromagnetically to their corresponding sublattice in the 
second layer (e.g. A-a), stabilized by the FM $\mathcal{J}_{Aa}$ coupling. The second group of solutions 
involves the Fe$_3X$ compounds, which display three degenerate FM solutions, as depicted in 
Fig.~\ref{Fig_CMP_energies}e (Nmltp=2,7,8). The ferromagnetic order within each Kagome layer is stabilized 
by the large FM $\mathcal{J}_{AB}$, while the FM order between the layers is stabilized by the strong FM 
$\mathcal{J}_{Ab}$, despite the small negative $\mathcal{J}_{Aa}$, as shown in Table~\ref{Tab_all_sub_int}.
\begin{figure*}
\centering
 \includegraphics[width=1.0\textwidth]{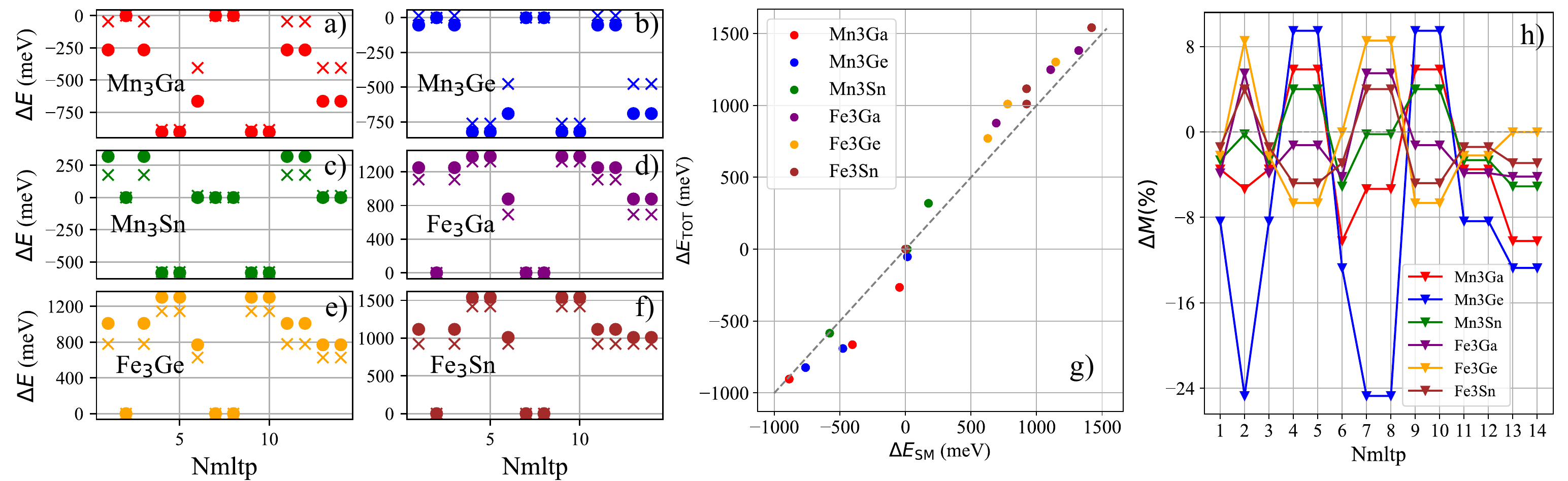}
  \caption{a–f) Comparison of the energy differences for all $Tm_3X$ compounds, obtained using the DLM spin model (crosses) and the total energy of constrained magnetic structures (circles), with the reference structure set to Nmltp=2 (ferromagnetic $\parallel$ $z$-axis). g) Deviation of the DLM spin model energies from the constrained structures; the dashed grey line indicates the $y=x$ curve. h) Deviation of the magnetic moments $(\Delta M)$ for the different constrained structures relative to the magnetic moment obtained from the charge self-consistent DLM calculations $(M_\text{DLM})$.}
\label{Constrains_vs_DLM}
\end{figure*}

The results obtained by including only the biquadratic terms in Eq.\eqref{sub_heisenberg} are shown 
in Fig.\ref{Fig_CMP_energies}b. This energy exhibits more degeneracies compared to Fig.~\ref{Fig_CMP_energies}a, 
where pair interactions are considered, as it includes terms of the form $(\VEC{e}_{n}\cdot\VEC{e}_{m})^2$, 
which do not distinguish between FM and AFM coupled moments. Instead, it splits the states into two subgroups: 
one comprising collinear states (Nmltp=2,6,7,8,13,14) and the other comprising non-collinear states 
(Nmltp=1,3,4,5,9,10,11,12), characterized by a $120^{\circ}$ relative angle within each Kagome layer. 
For Mn$_3X$, the biquadratic terms systematically favor a non-collinear order, whereas for Fe$_3X$, 
they favor a collinear order. This behavior is consistent with the predominantly negative (positive) 
$\mathcal{B}_{mn}$ coefficients for Mn$_3X$ (Fe$_3X$).

The energy of the CMP structures considering both pair and biquadratic terms is shown in 
Fig.~\ref{Fig_CMP_energies}c. The energy landscape remains similar to Fig.~\ref{Fig_CMP_energies}a, 
with modified energy barriers between collinear and non-collinear solutions due to the contributions 
from the biquadratic terms. The degeneracies between the CMP structures in Fig.~\ref{Fig_CMP_energies}a 
are further lifted by introducing relativistic chiral and non-chiral interactions~\cite{Sandratskii1996,park2018,Nyari2019}, 
which will be discussed in Sec.~\ref{relativistic_interactions}.  Figure~\ref{Fig_CMP_energies}d 
focuses on two compounds: Mn$_3$Sn (green curves) and Fe$_3$Sn (brown curves). Each compound has 
three curves: ``Full'', ``S1'', and ``S2''. Full: indicates that the energy is calculated by including 
all interactions between the sublattices. S1: includes only interactions within the same sublattice group, 
either $\{$A, B, C$\}$ or 
$\{$a, b, c$\}$. These interactions stabilize non-collinear CMP solutions in Mn$_3$Sn and collinear 
ones in Fe$_3$Sn. S2: includes only interactions between sublattices $\{$A, B, C$\}$ and $\{$a, b, c$\}$, 
which are crucial for two reasons: (i) they lift the degeneracy between AFM and FM coupled moments among 
two the Kagome layers (e.g., A–a and B–b), and (ii) in Mn$_3$Sn, they produce the largest energy difference 
between the low-energy solutions (Nmltp=4,5,9,10) and the remaining higher-energy ones.

To benchmark the energy of the CMP structures obtained from the DLM spin model (pair + biquadratic) against 
the total energy differences, we employ a dense 40$\times$40$\times$40 $k$-mesh and self-consistently
compute the total energies of the 14 CMP structures for each $Tm_3X$ compound, accounting for a total of 84 
calculations. The comparison of the energy differences $\Delta E$, taking structure Nmltp$=2$ (ferromagnetic 
$\parallel$ $z$-axis) as a reference is depicted in 
Figs.~\ref{Constrains_vs_DLM}a-f. To better illustrate the deviation between the spin model and the total energy 
calculations, we plot in Fig.~\ref{Constrains_vs_DLM}g the energy difference $\Delta E_\text{SM}$, obtained using 
Eq.~\eqref{sub_heisenberg}, on the $x$-axis, and $\Delta E_\text{TOT}$ on the $y$-axis, which denotes 
the total energy differences obtained directly from DFT. The deviations arise from several factors, 
including: (i) differences between the high-temperature DLM electronic structure and the zero-temperature 
one obtained in the constrained calculations, (ii) the omission of multi-site higher-order magnetic interactions, 
(iii) the contribution of the induced magnetic moments on the $X$ atoms, and (iv) variations in the Mn and Fe spin moment 
lengths among the constrained CMP structures. Despite these differences the DLM spin model successfully 
reproduces the global energy minima and the overall energy landscape of the CMP structures in comparison 
with the total energy calculations, at a significantly lower computational cost—i.e., one self-consistent 
DLM calculation versus 14 constrained ones for each system.

For each CMP structure, the spin moments have the same length $M$ at every site; however, this length 
varies depending on the structure. Fig.~\ref{Constrains_vs_DLM}h shows the deviation $\Delta M_\text{Nmltp} = (M_\text{Nmltp} - M_\text{DLM}) / M_\text{DLM}$ of the $Tm$ spin moments in the constrained CMP structures $M_\text{Nmltp}$ (Nmltp$=1,\dots,14$) relative to the DLM spin moment $(M_\text{DLM})$ given in Table~\ref{lattice_magn_1}. For Mn$_3X$, $\Delta M_\text{Nmltp}$ exhibits a positive deviation for the lowest-energy CMP structures (Nmltp$=4,5,9,10$), with values of $4\%$ for Mn$_3$Sn and $6\%$ for Mn$_3$Ga, while the largest deviation is 
found for Mn$_3$Ge at $9.5\%$. These deviations become significant, reaching $\Delta M_\text{Nmltp}=-24\%$, when 
the non-collinear Mn$_3$Ge is constrained to a collinear configuration (Nmltp$=2,7,8$). For Fe$_3X$ compounds, 
the corresponding low-energy CMP structures (Nmltp$=2,7,8$) exhibit a positive deviation $\Delta M_\text{Nmltp}$ 
of $4\%-8\%$, whereas imposing non-collinear structures (Nmltp$=4,5,9,10$) systematically reduces their spin moments. These deviations underscore the challenge of selecting a reference state without prior knowledge of the lowest-energy CMP structure when constructing the spin model. This motivates our choice of the DLM reference state.

\subsection{Relativistic interactions}
\label{relativistic_interactions}
Lastly, we analyze the role of the spin-orbit interaction (SOI) on lifting degeneracies of the seven 
lowest-energy CMP structures depicted in Fig.~\ref{Fig_CMP_energies}e. Considering the same computational 
setup as in Sec.\ref{setup_details}, the SOI is included self-consistently~\cite{bauer2014} and rotational 
symmetry in spin space is broken which prevents mapping the paramagnetic DLM state onto a simple up/down 
binary pseudo-alloy prescription. A careful treatment requires a relativistic extension of the DLM 
approach~\cite{Staunton2004,staunton2006}, in which averaging over multiple local-moment orientations 
is performed to capture the vectorial nature of thermal fluctuations~\cite{staunton2006,Szunyogh2011}. 
While we do not incorporate such advanced treatment of the spin fluctuations, we adopt a simplified approach 
for relativistic DLM using a pseudo-alloy with six components, corresponding to magnetic moments along 
the Cartesian directions $\{\pm x, \pm y, \pm z\}$, to capture the anisotropic nature of these 
fluctuations~\cite{Bouaziz2023}.

On one hand, the large isotropic pair and biquadratic interactions are weakly affected by the SOI. 
These minor changes do not influence the energy landscape of the CMP structures shown in Fig.~\ref{Fig_CMP_energies}. 
On the other hand, the SOI generates single-ion and two-ion symmetric anisotropic interactions~\cite{Nyari2019,Zelenskiy2021} 
(SAI) as well as anti-symmetric Dzyaloshinskii-Moriya interactions (DMI)~\cite{Nyari2019,Zelenskiy2021}, which remove 
the degeneracy among the identified low-energy CMP structures. Given that these structures split into two 
groups—non-collinear (Mn$_3X$) and collinear (Fe$_3X$)—we examine the role of the DMI in selecting the 
chirality for Mn$_3X$ CMP structures and the role of the SAI in stabilizing the low-energy configuration 
in Fe$_3X$. 
\subsection{Non-collinear systems: Mn$_3X$}
\label{DM_Mn3X}
\begin{figure}
  \centering
  \includegraphics[width=0.47\textwidth]{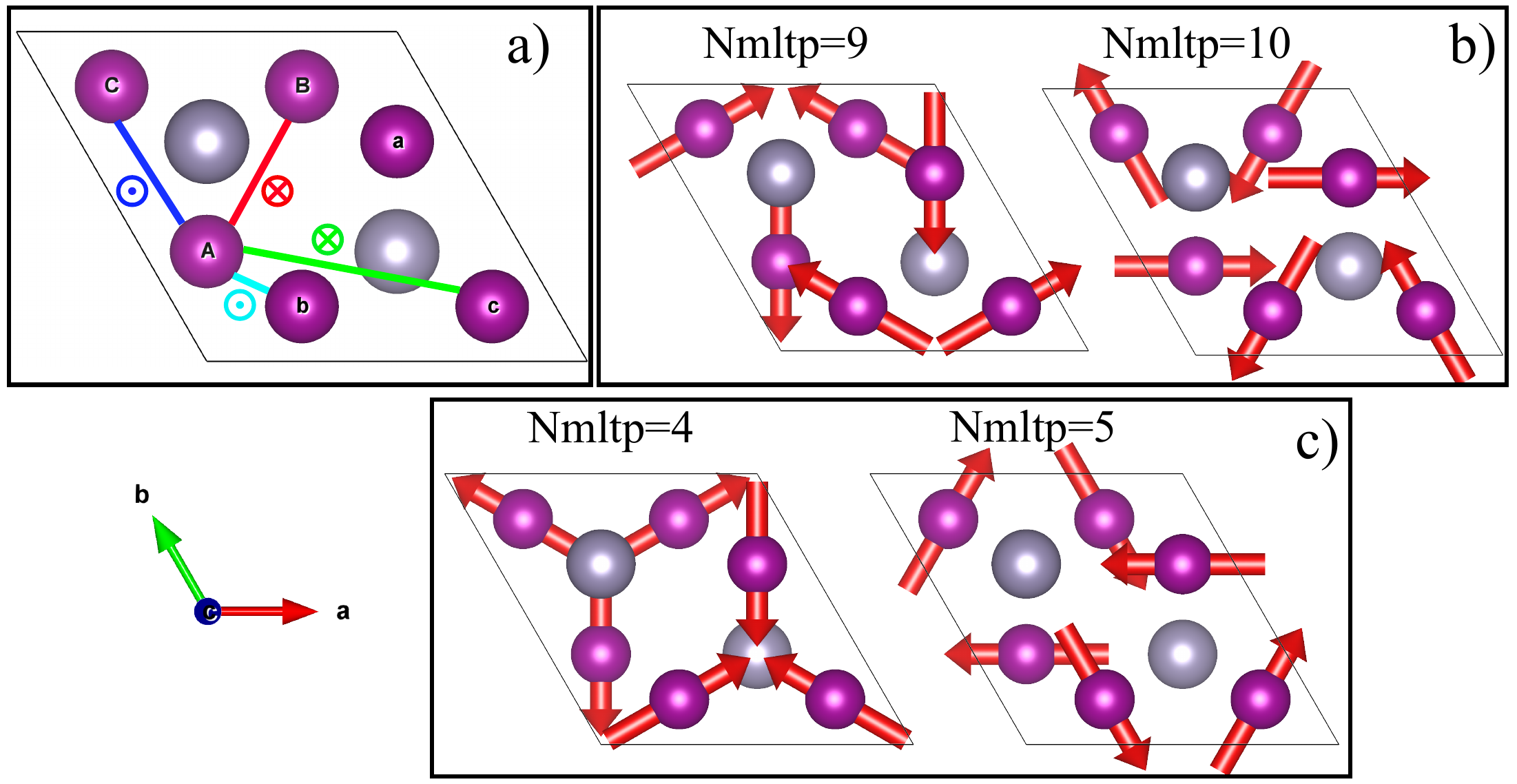}
\caption{a) Nonzero $z$-component of the sublattice Dzyaloshinskii-Moriya vector, 
indicated in different colors, starting from A: $\mathcal{D}_{AB}=-\mathcal{D}_{AC}$ 
and $\mathcal{D}_{Ab}=-\mathcal{D}_{Ac}$. b) Negative chirality (inverse triangular) 
structures that lower the chiral part of the Hamiltonian, $\mathcal{H}_{ch}$, given 
in Eq.~\eqref{heisen_DMI}. c) Positive chirality structures that are unfavored by 
$\mathcal{H}_{ch}$.}
\label{DMI_chiral_structures}
\end{figure}

As aforementioned, we inspect the contribution of the DMI to the non-collinear magnetic order in 
Mn$_3X$, focusing on the four lowest-energy non-collinear structures in Fig.~\ref{Fig_CMP_energies}e. 
Since these structures are coplanar in the $ab$-plane and their cross product points along the $c$-axis 
($z$-axis), only the $z$-component of the Dzyaloshinskii-Moriya vector $D^{z}_{ij}$ contributes to the energy.  
$D^{z}_{ij}$ is extracted via $D^{z}_{ij}=(\mathcal{J}^{xy}_{ij}-\mathcal{J}^{yx}_{ij})/2$ from the 
anisotropic pair exchange tensor components $\mathcal{J}^{\alpha\beta}_{ij}$ obtained using equation~\eqref{Full_Jij_appendix}, 
considering the averaged Green function $\bar{{\boldsymbol{G}}}_{ij}(E) = \bar{{\boldsymbol{G}}}^{+z+z}_{ij}(E)$. 
Similarly to the pair and biquadratic isotropic interactions, we introduce a sub-lattice model for the chiral 
part of the cell Hamiltonian $\mathcal{H}_{ch}$, which reads:
\begin{equation}
\begin{split}
\mathcal{H}_{ch} &= -\sum_{mn} \mathcal{D}^{z}_{mn}\,(\VEC{e}_{m}\times\VEC{e}_{n})^z\quad.\\ 
\end{split}
\label{Heisenberg_DMI}
\end{equation}
The summation ${m,n}$ in the previous equations runs over all sublattices $\{$A, B, C, a, b, c$\}$. 
The sublattice DMI coefficients are given by $\mathcal{D}^{z}_{mn}=\sum_{l} D^{z,0l}_{mn}$; these 
are antisymmetric and obey $\mathcal{D}^{z}_{mn}=-\mathcal{D}^{z}_{nm}$, with the same sublattice 
contributions canceling out ($\mathcal{D}^{z}_{nn}=0$). The computed DMI coefficients follow the 
symmetry requirements imposed by the D$_{6h}$ point group, the details of the symmetry analysis are 
given in Ref.~\onlinecite{Zelenskiy2021}, which takes into account the DMI interaction within each 
kagome plane and the interlayer couplings. These requirements impose several restrictions on the DMI 
vectors, allowing for a considerable simplification of the form of Eq.~\eqref{Heisenberg_DMI}, reducing 
the model to only two coefficients: $\mathcal{D}_{1}$ for interlayer and $\mathcal{D}_{2}$ for intralayer 
sublattice couplings. The DMI's relating inversion-symmetric sublattices vanish, leading to $\mathcal{D}^{z}_{Aa}=\mathcal{D}^{z}_{Bb}=\mathcal{D}^{z}_{Cc}=0$. The intra-layer couplings are related 
via $\mathcal{D}_{2}=\mathcal{D}^{z}_{AB}=-\mathcal{D}_{AC}=\mathcal{D}^{z}_{BC}$ (for $z=c/4$) and 
$\mathcal{D}_{2}=\mathcal{D}^{z}_{ab}=-\mathcal{D}^{z}_{ac}=\mathcal{D}^{z}_{bc}$ (for $z=3c/4$), 
while the interlayer ones are related via $\mathcal{D}_{1}=\mathcal{D}^{z}_{Ab}=-\mathcal{D}^{z}_{Ac}=\mathcal{D}^{z}_{aB}=-\mathcal{D}^{z}_{aC}=\mathcal{D}^{z}_{Bc}=\mathcal{D}^{z}_{bC}$. The chiral pair Hamiltonian simplifies 
as~\cite{Zelenskiy2021}:
\begin{equation}
\begin{split}
\mathcal{H}_{ch} &= -2 \mathcal{D}_{2}\left[
(\VEC{e}_{A}\times\VEC{e}_{B})^z
-(\VEC{e}_{A}\times\VEC{e}_{C})^z+
(\VEC{e}_{B}\times\VEC{e}_{C})^z\right.\\
&\left.+(\VEC{e}_{a}\times\VEC{e}_{b})^z
-(\VEC{e}_{a}\times\VEC{e}_{c})^z+
(\VEC{e}_{b}\times\VEC{e}_{c})^z\right]\\
&
-2 \mathcal{D}_{1}
\left[
 (\VEC{e}_{A}\times\VEC{e}_{b})^z
-(\VEC{e}_{A}\times\VEC{e}_{c})^z-
 (\VEC{e}_{B}\times\VEC{e}_{a})^z\right.\\
&\left.
+(\VEC{e}_{C}\times\VEC{e}_{a})^z+
 (\VEC{e}_{B}\times\VEC{e}_{c})^z-
 (\VEC{e}_{C}\times\VEC{e}_{b})^z\right]\,.\\
\end{split}
\label{heisen_DMI}
\end{equation}
The first two lines represent the contribution of the two Kagome layers, while the last 
two represents the interlayer coupling. The values of $\mathcal{D}_{1}$ and $\mathcal{D}_{2}$ 
for the Mn$_3X$ compounds are given in Table~\ref{DMI_table}. The highest values for 
$\mathcal{D}_{1}$ and $\mathcal{D}_{2}$ are obtained for Mn$_3$Sn, followed by Mn$_3$Ge, 
and the smallest for Mn$_3$Ga. The chiral energy contribution $\Delta\mathcal{E}_{ch}$ 
is obtained from Eq.~\eqref{heisen_DMI} using the magnetic structures given in Fig.~\ref{DMI_chiral_structures}b). 
These correspond to the inverse triangular structures discussed in Refs.~\onlinecite{Sandratskii1996,Soh2020,Zelenskiy2021}, 
while the structures with opposite chirality (Fig.~\ref{DMI_chiral_structures}c) are unstable 
with the opposite energy $-\Delta\mathcal{E}_{ch}$. This chiral energy contribution is two orders 
of magnitude smaller than the previously discussed isotropic pair and biquadratic energies, 
as it originates from the SOI~\cite{bouaziz2017}. Moreover, the interlayer and intralayer 
sublattice couplings are in competition, as they have opposite signs for all Mn$_3X$ compounds. 
The negative chirality is fixed by $\mathcal{D}_{2}$. The chiral energy contribution, setting 
$\mathcal{D}_{1}$ to zero $(\Delta\mathcal{E}^\text{NL}_{ch})$, is larger in magnitude compared 
to $\Delta\mathcal{E}_{ch}$, and the negative chirality structures are further stabilized.

The combination of single and two-ion anisotropy lifts the degeneracy between the two inverse triangular structures. A detailed discussion of the SAI on the magnetic order of Mn$_3X$ is provided in earlier studies~\cite{Nyari2019,Zelenskiy2021}. These interactions 
were observed to be of weak magnitude (tens of $\mu$eV); nonetheless, they play an important role in further lifting the degeneracy among non-collinear CMP structures (Fig.~\ref{DMI_chiral_structures}b) and inducing net magnetic moment in the $ab$-plan (weak ferromagnetism)~\cite{Nyari2019,Zelenskiy2021}. Note that the analysis of weak ferromagnetism in Mn$_3X$ requires going beyond the spin model approach presented here to account for orbital polarization effects and induced moments on the $X$ sites~\cite{Sandratskii1996}.
\begin{table}
\centering
\begin{tabular}{ccccc}
\hline
System  & $\mathcal{D}_{1}$ & $\mathcal{D}_{2}$ & $\Delta\mathcal{E}_{ch}$ & $\Delta\mathcal{E}^\text{NL}_{ch}$ \\
\hline
Mn$_3$Ga & 0.18 & -0.31 &  -1.397  & -3.266\\
Mn$_3$Ge & 0.30 & -0.41 &  -1.088  & -4.231\\
Mn$_3$Sn & 0.41 & -0.65 &  -2.502  & -6.724\\
\hline
\end{tabular}
\caption{
Sublattice DMI coefficients $(\mathcal{D}_{1}, \mathcal{D}_{2})$ for Mn$_3X$. 
$\Delta\mathcal{E}_{ch}$ is the chiral energy contribution for the negative chiral 
structures (Fig.~\ref{DMI_chiral_structures}b). $\Delta\mathcal{E}^\text{NL}_{ch}$ 
is the same as $\Delta\mathcal{E}_{ch}$, with $\mathcal{D}_{1}$ set to zero. 
All values are given in meV.}
\label{DMI_table}
\end{table} 

\subsection{Collinear systems: Fe$_3X$}
\begin{figure}
  \centering
 \includegraphics[width=0.45\textwidth]{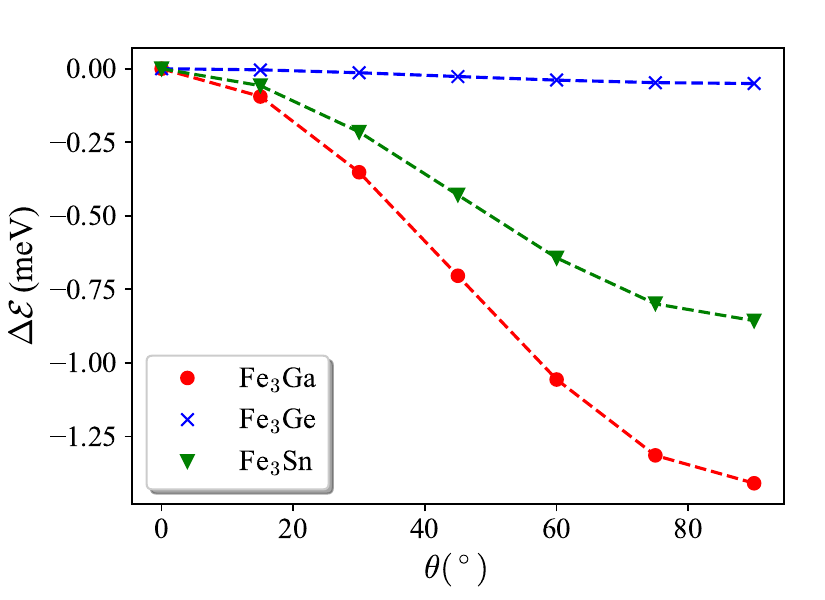}
  \caption{Magnetic anisotropy of the Fe$_3X$ as function of the polar angle $\theta$. 
  The dashed lines indicate the fits using the functional form given in Eq.~\eqref{pheno_mae}. 
  All the systems favor an in-plan orientation of the magnetic moments.}
\label{MAE_Fe3X_polar}
\end{figure}

The magnetic anisotropy energy changes as a function of temperature~\cite{skomski2008}, 
which is due to changes in the electronic structure as thermal spin fluctuations increase. 
The temperature dependence of the anisotropy energy can be extracted via the magnetic torque 
method~\cite{staunton2006}. It consists of single-ion and two-ion contributions, which should 
be distinguished when dealing with non-collinear magnetic orders~\cite{amoroso2020}. 
Focusing on the Fe$_3X$ which were determined to be ferromagnetic (Nmltp=2,7,8 from Fig.~\ref{Fig_CMP_energies}e) at zero temperature, the anisotropy energy can be expressed in 
terms of effective anisotropy constants $\mathcal{K}_{i}$, which are given by the sum of single-ion 
and two-ion contributions~\cite{staunton2006,Bouaziz2023}. 
These constants $\mathcal{K}_{i}$ are determined by fitting the self-consistent total energy variations of the unit cell under global rotations of $\VEC{m}_i$. The phenomenological form of the anisotropy energy $\Delta \mathcal{E}$ for the hexagonal uniaxial 
Fe$_3X$ systems at hand is given by~\cite{skomski2008}:
\begin{equation}
\begin{split}
\Delta \mathcal{E}(\theta,\phi) & = \mathcal{K}_{1}\sin^2\theta + \mathcal{K}_{2}\sin^4\theta \\
& + \mathcal{K}_{3}\sin^6\theta +
\mathcal{K}^{\prime}_{3}\sin^6\theta\cos6\phi\quad,    
\end{split}
\label{pheno_mae}
\end{equation}
where $\theta$ is the polar angle between the magnetic moment and the $c$-axis, and $\phi$ is 
the azimuthal angle between the moment and the $a$-axis. The values of $\Delta \mathcal{E}$ obtained 
for the Fe$_3X$ as a function of $\theta$ are shown in Fig.~\ref{MAE_Fe3X_polar}, while the fitted 
values of the constants $\mathcal{K}_{1}$ and $\mathcal{K}_{2}$ are given in Table~\ref{MAE_Fe3X_table}. 
All the Fe$_3X$ favor an in-plane orientation of the magnetic moments $(\theta=90^\circ)$. The largest 
anisotropy barrier is obtained for Fe$_3$Ga, intermediate for Fe$_3$Sn, and the lowest for Fe$_3$Ge. 
This lifts the energy degeneracy, favoring the in-plane CMP structures Nmltp=7,8 over the out-of-plane 
structure Nmltp=2. $\Delta \mathcal{E}$ is dominated by second-order constants with 
$\mathcal{K}{2} \ll \mathcal{K}_{1}$ and $\mathcal{K}_{3}\simeq 0$. The sixth-order basal anisotropy 
is extremely weak, and its constant $\mathcal{K}^{\prime}_{3}$ is of the order of $\simeq 1\mu$eV. 
Thus, in-plane CMP structures Nmltp=7,8 are quasi-degenerate. Other effects, such as small lattice 
distortions or doping, can contribute to lifting this degeneracy.
\begin{table}
\centering
\begin{tabular}{ccc}
\hline
System  & $\mathcal{K}_{1}$ & $\mathcal{K}_{2}$ \\
\hline
Fe$_3$Ga & -1.407 & -0.002  \\
Fe$_3$Ge & -0.056 &  0.005  \\
Fe$_3$Sn & -0.860 &  0.004 \\
\hline
\end{tabular}
\caption{Effective magnetic anisotropy constants $\mathcal{K}_{1}$ and $\mathcal{K}_{2}$ 
for the hexagonal uniaxial Fe$_3X$ compounds. All the values are given in meV.}
\label{MAE_Fe3X_table}
\end{table}

\section{Conclusions}
\label{All_conclusions}
We introduced a multi-scale computational approach which relies on a robust atomistic spin model parameters using the paramagnetic state as a reference, targeting both collinear and non-collinear magnetic systems. The orthogonal and complete basis of magnetic structures was systematically generated using the cluster multipole (CMP) method, bypassing the need for additional atomistic spin dynamics or Monte Carlo simulations. The $Tm_3X$ kagome compounds provided an ideal platform to apply this approach, successfully recovering the magnetic structures reported from previous experimental and theoretical works~\cite{Nyari2019,sticht1989,Mendive2019,Soh2020,nakatsuji2015,Sales2014,Zheng2}. 

The approach offers several advantages, as demonstrated for the complex $Tm_3X$ kagome compounds, where it enables: (i) the inspection of arbitrary magnetic configurations while maintaining constant-moment solutions at minimal computational cost, (ii) the identification of commensurate or incommensurate magnetic order and a rapid estimation of the transition temperature within mean-field theory, (iii) the analysis of the role of pair and higher-order interactions in selecting low-energy CMP structures, leading to the identification of a significant biquadratic interaction contribution in $Tm_3X$—promoting non-collinear CMP structures in Mn$_3X$ and collinear CMP structures in Fe$_3X$, (iv) the disentangling of multiple sublattice interactions, which is crucial for unit cells containing multiple atoms, as exemplified by the Mn$_3X$ compounds, where we found competing intralayer and interlayer Dzyaloshinskii–Moriya interactions (DMI), and (v) the incorporation of smaller energy-scale interactions arising from relativistic effects, including the antisymmetric DMI, which determines the chirality of the CMP structures, as well as single-ion and two-ion anisotropies, which further lift the degeneracy among the identified low-energy CMP structures.

\textit{Outlook:} In this work, we focused on compounds with well-characterized magnetic structures, both theoretically and experimentally. The proposed framework provides an efficient approach for exploring material families with unknown magnetic structures by performing high-throughput computations of spin model parameters from the paramagnetic (DLM) state, combined with the CMP method, to systematically identify low-energy magnetic configurations. The low computational cost of energy evaluations at the model level enables the study of more complex cases, including magnetic order derived from linear combinations of CMP structures~\cite{Huebsch2021}, as well as systems with large unit cells or incommensurate magnetic orders~\cite{Bouaziz2022,Bouaziz2023} using a recently developed CMP extension~\cite{Yanagi2023}. Moreover, extending spin models beyond two-site interactions to incorporate higher-order multi-site interactions, such as isotropic four-spin interactions, allows for a more accurate description of exotic magnetic phases, including skyrmion lattices~\cite{heinze2011} and meronic phases~\cite{khanh2022}.  

\section*{Acknowledgments} 
J. B. thanks Phivos Mavropoulos for fruitful discussions. J. B. was supported by the Alexander von Humboldt Foundation through the Feodor Lynen Research Fellowship for Postdocs. This work was supported by RIKEN Junior Research Associate Program. This work was supported by the RIKEN TRIP initiative (RIKEN Quantum, Advanced General Intelligence for Science Program, Many-body Electron Systems). We acknowledge the financial support by Grant-in-Aids for Scientific Research (JSPS KAKENHI) Grant Numbers JP21H04990, JP22H00290, and JP24K00581, JST-CREST No.~JPMJCR23O4, JST-ASPIRE No.~JPMJAP2317, JST-Mirai No.~JPMJMI20A1.
\appendix

\section{Magnetic force theorem for ferromagnetic state}
\label{force_theorem_FM}
In this appendix, we recall the mapping of the magnetic interactions from the ferromagnetically ordered state, including pair~\cite{Liechtenstein1984,Udvardi2003,Ebert2009} and the extension to higher order biquadratic interactions\cite{Mankovsky2020,Lounis2020}. The effect of spin-orbit interaction are also considered, therefore, allowing the emergence of isotropic/anisotropic pair and higher order terms in the spin Hamiltonian. The variation of the band energy is due to an infinitesimal rotation in the magnetic exchange-correlation potential $(B^\text{xc}(\vec{r}))$ to different orders can be easily extracted from the Lloyds formula~\cite{lloyd1972,drittler1989}:
\begin{equation}
\begin{split}
\delta\mathcal{U} & = -\frac{1}{\pi}\text{Im}\text{Tr}_{i\sigma}\int_{-\infty}^{E_F}
dE\,\sum_{n}\frac{1}{n}(\boldsymbol{\mathcal{G}}(E)\delta \boldsymbol{V})^{n},
\end{split}
\label{lloyd_form}
\end{equation}
where $\mathcal{U}$ is the band energy, $\boldsymbol{\mathcal{G}}(E)$ is the Green function of the 
system, $\delta\boldsymbol{V}$ is the variation of potential due the infinitesimal 
rotation. The trace is taken over spin $(\sigma)$ and sites $(i)$. The variation of the 
band energy due to a two-site perturbation in the system is obtained by expanding 
Eq.~\eqref{lloyd_form} to the second order: 
\begin{equation}
\begin{split}
\delta\mathcal{U}_{2} & = -\frac{1}{2\pi}\text{Im}\text{Tr}_{\sigma}\sum_{i\ne j}
\int_{-\infty}^{E_F}dE\,\boldsymbol{\mathcal{G}}_{ij}(E)\delta \boldsymbol{V}_j\,\boldsymbol{\mathcal{G}}_{ji}(E)\delta \boldsymbol{V}_i\,.
\label{two_sites}
\end{split}
\end{equation}
The single particle Kohn-Sham potential at a given site $i$, $\boldsymbol{V}_{i}(\vec{r})$ can be splitted into 
a charge part ${V}_{i}^{0}(\vec{r})$ and a magnetic contribution $\vec{B}_{i}^\text{xc}(\vec{r})$: 
$V_{i}(\vec{r})=V_{i}^{0}(\vec{r})\,\boldsymbol{\sigma}_{0}+\vec{B}^\text{xc}_{i}(\vec{r})
\cdot\vec{\boldsymbol{\sigma}}$, where $\boldsymbol{\sigma}_{0}$ is $2\times2$ unit matrix and $\vec{\boldsymbol{\sigma}}=\{\boldsymbol{\sigma}_x,\boldsymbol{\sigma}_y,\boldsymbol{\sigma}_z\}$ is the Pauli vector. In the rigid spin approximation, the variation of the potential reads~\cite{phariseau2012}: 
\begin{equation}
\delta \boldsymbol{V}_{i}(\vec{r}\,) = \sum_{\alpha=x,y,z}
B^\text{xc}_{i}(\vec{r}\,)\,(\delta{e}^{\alpha}_{i}
\cdot{\boldsymbol{\sigma}}^{\alpha})\quad.
\label{delta_vi}
\end{equation}
We now introduce the orbital expansion explictely in the KKR representation of the Green function~\cite{Ebert2009,Ebert2011,Papanikolaou2002,bauer2014} as:
\begin{equation}
\begin{split}
\boldsymbol{\mathcal{G}}_{ij}&(\vec{r},\vec{r}^{\,\prime},E)  = \boldsymbol{G}^\text{o}_{ii}(\vec{r},\vec{r}^{\,\prime},E)\,\delta_{ij} 
\\& + 
\sum_{LL^{\prime}}\boldsymbol{R}_{iL}(\vec{r},E)\,\boldsymbol{G}_{ij,LL^{\prime}}(E)
\,\boldsymbol{\bar{R}}_{jL^{\prime}}(\vec{r}^{\,\prime},E),
\end{split}
\label{kkr_gf_full}
\end{equation}
$L=(l,m)$ is the angular orbital momentum expansion index, $\boldsymbol{G}^\text{o}_{ii}(\vec{r},\vec{r}^{\,\prime},E)$
is the on-site Green function which does not contribute to the magnetic interactions since $i \ne j$, 
$\boldsymbol{G}_{ij,LL^{\prime}}(E)$ is the structural Green function, and $\boldsymbol{R}_{iL}(\vec{r},E)$ 
($\boldsymbol{\bar{R}}_{iL}(\vec{r},E)$) is the right (left) regular scattering solutions. Inserting the KKR 
representation of the Green function (Eq.~\eqref{kkr_gf_full}) and the explicit form $\delta \boldsymbol{V}_i(\vec{r})$
(Eq.~\eqref{delta_vi}) into Eq.~\eqref{two_sites} $\delta\mathcal{U}_{2}$ is expressed as:
\begin{equation}
\begin{split}
\delta\mathcal{U}_{2} 
& = -\frac{1}{2\pi}\,\text{Im}\,\text{Tr}_{\sigma,L}\sum_{i\neq j} 
\sum_{\alpha\beta}\int^{E_\text{F}}_{-\infty} dE
\\&
\,{\boldsymbol{G}}_{ij}(E) 
\,{\boldsymbol{B}}^{\beta}_{j}(E)
\,{\boldsymbol{G}}_{ji}(E) 
\,{\boldsymbol{B}}^{\alpha}_{i}(E)
\, \delta{e}^\alpha_{i} \delta{e}^\beta_{j}\quad.\\& 
\end{split}
\label{final_energy_kkr}
\end{equation}
Note that the bold quantities in the previous equations are matrices in spin $(\sigma)$ and the angular momentum $L$ space and the trace is taken accordingly. The sum over $\{\alpha,\beta\}$ runs over the three spatial directions $\{x,y,z\}$. The matrix ${\boldsymbol{B}}^{\alpha}_{i}(E)$ as introduced as~\cite{Ebert2009}:
\begin{equation}
\begin{split}
\boldsymbol{B}^{\alpha}_{iL_1L_2}(E) &= \int d\vec{r}\,\boldsymbol{\bar{R}}_{iL_1}(\vec{r},E)
\left[B^\text{xc}_{i}(\vec{r})\,{\boldsymbol{\sigma}}_{\alpha}\right]\boldsymbol{R}_{iL_2}(\vec{r},E)\quad.
\end{split}
\label{tmat_ebert}
\end{equation}
The analytical expression for the pair magnetic exchange interaction tensor is then obtained by mapping the internal 
energy variation $\delta\mathcal{U}_{2}$ (Eq.~\eqref{final_energy_kkr}) onto the following pair anisotropic Heisenberg 
Hamiltonian: $\mathcal{H}_\text{}= -\sum_{i\neq j}\sum_{\alpha\beta}\mathcal{J}^{\alpha\beta}_{ij}\,{e}^\alpha_{i}\,{e}^\beta_{j}$
leading to the expression~\cite{Ebert2009,bauer2014}:
\begin{equation}
\begin{split}
\mathcal{J}^{\alpha\beta}_{ij} & = 
\frac{1}{2\pi}\,\text{Im}\,\text{Tr}_{\sigma,L}
\int^{E_\text{F}}_{-\infty} dE \\&
\,{\boldsymbol{G}}_{ij}(E) 
\,{\boldsymbol{B}}^{\beta}_{j}(E)
\,{\boldsymbol{G}}_{ji}(E) 
\,{\boldsymbol{B}}^{\alpha}_{i}(E)\quad.
\end{split}
\label{Full_Jij_appendix}
\end{equation}
The higher order interactions, in particular the biquadratic interactions are extracted from 
the $4^\text{th}$ order expansion of Lloyds formula (Eq.~\eqref{lloyd_form}):
\begin{equation}
\begin{split}
\delta\mathcal{U}_{4} & = -\frac{1}{4\pi}\text{Im}\text{Tr}_{\sigma}\sum_{ijkl}
\int_{-\infty}^{E_F}dE\\&
\boldsymbol{\mathcal{G}}_{ij}(E)
\delta \boldsymbol{V}_j\,
\boldsymbol{\mathcal{G}}_{jk}(E)
\delta\boldsymbol{V}_k\,
\boldsymbol{\mathcal{G}}_{kl}(E)
\delta \boldsymbol{V}_l\,
\boldsymbol{\mathcal{G}}_{li}(E)
\delta \boldsymbol{V}_i\quad.\\
\end{split}
\label{fourth_order_lloyds}
\end{equation}
Focusing the biquadratic interaction which involve only two sites, we consider the following case $k=i$, $l=j$ and $i\neq j$:
\begin{equation}
\begin{split}
\delta\mathcal{U}^\text{Biq}_{4} & = 
-\frac{1}{2\pi}\text{Im}\text{Tr}_{\sigma}\sum_{i\ne j}
\int_{-\infty}^{E_F}dE\\&
\,\boldsymbol{\mathcal{G}}_{ij}(E)
\delta \boldsymbol{V}_j\,
\boldsymbol{\mathcal{G}}_{ji}(E)
\delta \boldsymbol{V}_i\,
\boldsymbol{\mathcal{G}}_{ij}(E)
\delta \boldsymbol{V}_j\,
\boldsymbol{\mathcal{G}}_{ji}(E)
\delta \boldsymbol{V}_i\quad.
\end{split}
\end{equation}
Note that Eq.~\eqref{fourth_order_lloyds} involves a double summation over $\{i,j\}$ as only two sites are considered but the interaction is of fourth order. This results in extra factor two in the above expression. Introducing the explicit form of $\delta \boldsymbol{V}_i$ using Eq.~\eqref{delta_vi} and the Green function's expression in the KKR basis (Eq.~\eqref{kkr_gf_full}) results in:
\begin{equation}
\begin{split}  
\delta\mathcal{E}^\text{Biq}_{4} &
= -\frac{1}{2\pi}\,\text{Im}\,\text{Tr}_{\sigma,L}\sum_{i\neq j} 
\sum_{\alpha\beta\gamma\delta}\int^{E_\text{F}}_{-\infty} dE\\&
[\,{\boldsymbol{G}}_{ij}(E) 
\,{\boldsymbol{B}}^{\beta}_{j}(E)
\,{\boldsymbol{G}}_{ji}(E) 
\,{\boldsymbol{B}}^{\alpha}_{i}(E)]\\&
[\,{\boldsymbol{G}}_{ij}(E) 
\,{\boldsymbol{B}}^{\gamma}_{j}(E)
\,{\boldsymbol{G}}_{ji}(E) 
\,{\boldsymbol{B}}^{\delta}_{i}(E)]
\quad\\& \times \delta{e}^\alpha_{i} \delta{e}^\beta_{j}
\delta{e}^\gamma_{i} \delta{e}^\delta_{j} 
\end{split}
\label{final_energy_4th}
\end{equation}
The mapping is then done considering an anisotropic biquadratic Hamiltonian~\cite{Brinker2019}:
\begin{equation}
\begin{split}
\mathcal{H}_\text{biq} &= -\sum_{i\ne j}\sum_{\alpha\beta\gamma\delta} \mathcal{B}^{\alpha\beta\gamma\delta}_{ij}
{e}^{\alpha}_{i}{e}^{\beta}_{j}{e}^\gamma_{i}{e}^\delta_{j} \quad.     
\end{split}
\label{H_biq}
\end{equation}
A one to one identification between Eq.~\eqref{final_energy_4th} and Eq.~\eqref{H_biq} provides the analytical expression
of the biquadratic tensor as~\cite{Mankovsky2020}: 
\begin{equation}
\begin{split}
\mathcal{B}^{\alpha\beta\gamma\delta}_{ij} &= 
\frac{1}{2\pi}\,\text{Im}\,\text{Tr}_{\sigma,L}
\int^{E_\text{F}}_{-\infty} dE\\&
[\,{\boldsymbol{G}}_{ij}(E) 
\,{\boldsymbol{B}}^{\beta}_{j}(E)
\,{\boldsymbol{G}}_{ji}(E) 
\,{\boldsymbol{B}}^{\alpha}_{i}(E)]\\&
[\,{\boldsymbol{G}}_{ij}(E) 
\,{\boldsymbol{B}}^{\gamma}_{j}(E)
\,{\boldsymbol{G}}_{ji}(E) 
\,{\boldsymbol{B}}^{\delta}_{i}(E)]
\quad.
\end{split}
\label{biquadratic_ferro}
\end{equation}
The anisotropic biquadratic tensor is of fourth rank, nonetheless, the form of $\mathcal{H}_\text{biq}$ (Eq.~\eqref{H_biq}) provides some symmetry constrains on its components. Interchanging $\alpha$ ($\beta$) and $\gamma$ ($\delta$) lets the Hamiltonian invariant hence one can deduce the following symmetry 
properties~\cite{Brinker2019}: $\mathcal{B}^{\alpha\beta\gamma\delta}_{ij} = \mathcal{B}^{\gamma\beta\alpha\delta}_{ij}$, 
$\mathcal{B}^{\alpha\beta\gamma\delta}_{ij}  = \mathcal{B}^{\alpha\delta\gamma\beta}_{ij}$, 
$\mathcal{B}^{\alpha\beta\gamma\delta}_{ij} = \mathcal{B}^{\gamma\delta\alpha\beta}_{ij}$. 
\section{Necessary isotropic and anisotropic interactions for $Tm_3X$ compounds}
\label{parameters_extraction}
The pair exchange tensor containing the $\mathcal{J}^{\alpha\beta}_{ij}$ elements can be decomposed isotropic, anti-symmetric and anisotropic symmetric parts~\cite{Udvardi2003,Ebert2009,bouaziz2017,Mankovsky2020}. Considering that the magnetic moments in the ferromagnetic state point in the $z$-direction, the transverse part of the pair Hamiltonian accessible through the infinitesimal rotation approach can be conveniently re-expressed as:
\begin{equation}
\begin{split}
&\mathcal{H}^\text{T} = -\sum_{ij} J_{ij}\,({e}^x_{i}{e}^x_{j}+{e}^y_{i}{e}^y_{j})-\sum_{ij} D^{z}_{ij}\,(\VEC{e}_{i}\times\VEC{e}_{j})^{z}\\
& -\sum_{ij} S_{ij} (e^{x}_{i}e^{x}_{j}-e^{y}_{i}e^{y}_{j})
 -\sum_{ij} A_{ij} (e^{x}_{i}e^{y}_{j}+e^{y}_{i}e^{x}_{j})\,.
\end{split}
\end{equation}
The quantities introduced are the isotropic pair interaction $J_{ij}$, the $z$-component of the Dzyaloshinskii–Moriya vector $(D^{z}_{ij})$ and the two-ion anisotropy constants $S_{ij}$ and $A_{ij}$. These are the central quantities of interest discussed in the main text. These constants are related to components of the pair exchange tensor via:
\begin{equation}
\begin{split}
J_{ij} &= \frac{\mathcal{J}^{xx}_{ij}   + \mathcal{J}^{yy}_{ij}}{2}\,,\,
D^z_{ij} = \frac{\mathcal{J}^{xy}_{ij} - \mathcal{J}^{yx}_{ij}}{2}\,,\,\\ 
S_{ij} &= \frac{\mathcal{J}^{xx}_{ij}   - \mathcal{J}^{yy}_{ij}}{2}\,,\,
A_{ij} = \frac{\mathcal{J}^{xy}_{ij}  + \mathcal{J}^{yx}_{ij}}{2}\,.\\    
\end{split}
\end{equation}
In the scalar relativistic limit, \textit{i.e.} in absence of the spin-orbit interactions, the constants  $D^z_{ij},S_{ij}$ and $A_{ij}$ vanish~\cite{bouaziz2017}, and the isotropic pair interaction simplifies to:
\begin{equation}
\begin{split}
J_{ij} & = 
\frac{1}{2\pi}\,\text{Im}\,\text{Tr}_{\sigma,L}
\int^{E_\text{F}}_{-\infty} dE \\&
\,{\boldsymbol{G}}_{ij}(E) 
\,{\boldsymbol{B}}^{x}_{j}(E)
\,{\boldsymbol{G}}_{ji}(E) 
\,{\boldsymbol{B}}^{x}_{i}(E)\quad.
\end{split}
\label{Append_FM_eq_1}
\end{equation}

The biquadratic isotropic interactions can be extracted from the full tensor $\underline{\mathcal{B}}_{ij}$ similarly to the pair ones. Considering the transverse anisotropic symmetric part of the biquadratic Hamiltonian defined as:
\begin{equation}
\begin{split}
\mathcal{H}^\text{T}_\text{biq}&= -\sum_{i\ne j} \Big{[}B^{xxxx}_{ij}e^x_ie^x_je^x_ie^x_j+B^{xxyy}_{ij}
e^x_ie^x_je^y_ie^y_j\\&+B^{yyxx}_{ij}e^y_ie^y_je^x_ie^x_j+B^{yyyy}_{ij}e^y_ie^y_je^y_ie^y_j\Big{]}\quad.
\end{split}
\end{equation}
The latter can be re-expressed as:
\begin{equation}
\begin{split}
\mathcal{H}^\text{T}_\text{biq}&= -\sum_{i\ne j} B_{ij}
\Big{[}e^x_ie^x_je^x_ie^x_j+e^x_ie^x_je^y_ie^y_j\\
&+e^y_ie^y_je^x_ie^x_j+e^y_ie^y_je^y_ie^y_j\Big{]}+\mathcal{H}^\text{TA}_\text{biq}\quad.
\end{split}
\end{equation}
$\mathcal{H}^\text{TA}_\text{biq}$ contains two-ion biquadratic anisotropic contributions 
of the form $(B^{\alpha\alpha\beta\beta}_{ij}-B_{ij})\,e^\alpha_ie^\alpha_je^\beta_ie^\beta_j$. 
For periodic systems, these are generally weak, originating from a fourth-order scattering 
process and being second order in spin-orbit interaction. The isotropic biquadratic interactions 
is defined as~\cite{Mankovsky2020}:
\begin{equation}
B_{ij} = \frac{B^{xxxx}_{ij}+B^{xxyy}_{ij}+B^{yyxx}_{ij}+B^{yyyy}_{ij}}{4}\quad.
\end{equation}
Once more, in the scalar relativistic limit the anisotropic contributions vanish and 
$\mathcal{H}^\text{TA}_\text{biq}=0$. The $B_{ij}$ constant is then simply given by:
\begin{equation}
\begin{split}
{B}_{ij} &= 
\frac{1}{2\pi}\,\text{Im}\,\text{Tr}_{\sigma,L}
\int^{E_\text{F}}_{-\infty} dE\\&
[\,{\boldsymbol{G}}_{ij}(E) 
\,{\boldsymbol{B}}^{x}_{j}(E)
\,{\boldsymbol{G}}_{ji}(E) 
\,{\boldsymbol{B}}^{x}_{i}(E)]\\&
[\,{\boldsymbol{G}}_{ij}(E) 
\,{\boldsymbol{B}}^{x}_{j}(E)
\,{\boldsymbol{G}}_{ji}(E) 
\,{\boldsymbol{B}}^{x}_{i}(E)]
\quad.
\end{split}
\label{biquadratic_ferro}
\end{equation}
\begin{figure*}[hbt!]
	\centering
	\includegraphics[width=0.8\textwidth]{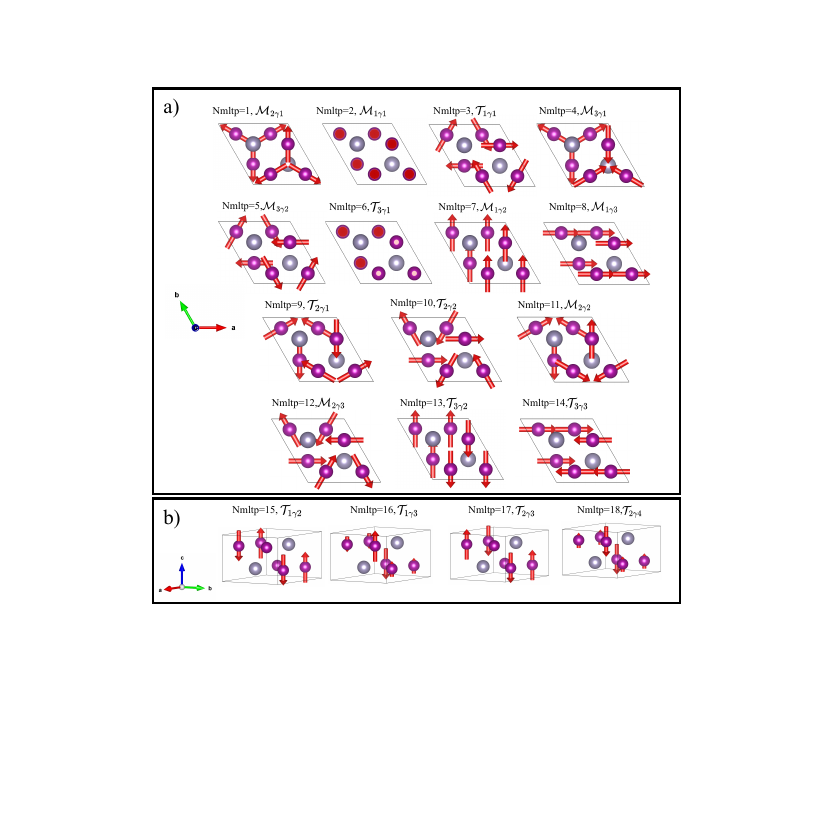}
    \caption{Generated CMP structures up to $l=3$.  
    (a) Magnetic ($\VEC{\mathcal{M}}_{l\gamma}$) and toroidal ($\VEC{\mathcal{T}}_{l\gamma}$) multipole structures, including collinear and non-collinear ones, that conserve the length of the magnetic moment $\VEC{m}_{i}$ are considered in this energy evaluation in Sec.~\ref{CMP_energetics}. (b) High-energy toroidal multipole structures involving large variations of the length of $\VEC{m}_{i}$.}
    \label{Full_MLTP}
\end{figure*}

\section{Complete CMP generated structures}
\label{Full_CMP_solutions}
The complete set of cluster multipole structures generated for $Tm_3X$ (space group P6$_3$/mmc) up to $l=3$ is depicted 
in Fig.~\ref{Full_MLTP}a. These structures are labeled with an integer index (Nmltp) and correspond to the respective 
magnetic ($\VEC{\mathcal{M}}_{l\gamma}$) and toroidal ($\VEC{\mathcal{T}}_{l\gamma}$) multipoles introduced in Eq.~\eqref{multipoles_def_1} 
of the main text. The simplest solutions are the dipole structures $\VEC{\mathcal{M}}_{1\gamma}$, which represent 
three collinear magnetic structures with moments $\VEC{m}_{i}$ oriented along the $(x,y,z)$ spatial directions (Nmltp$=2,7,8$). 
These structures account for spin-orbit anisotropy, as the CMP approach is based on the magnetic space group~\cite{Suzuki2019}. 
Fig.~\ref{Full_MLTP}b shows toroidal moment structures involving \textit{large} longitudinal variations in the magnitude 
of $\VEC{m}_{i}$. These structures involving drastic changes in the magnetic moment lengths are higher in energy and 
not considered in the evaluation of CMP energies in Sec.~\ref{CMP_energetics}. Our model analysis focuses on the 
low-energy configurations that involve directional changes of the magnetic moments while keeping their length fixed 
(transverse variations), as the $Tm$ atoms have local moments, for which the DLM approach provides a good description.  
 
\bibliography{bibliography.bib}

\end{document}